\documentclass[preprint]{aastex}

\newcommand{\kms}{km~s$^{-1}$}

\newcommand{\kcorrect}{{\ttfamily kcorrect~}}

\begin{document}

\title{Using the {\it XMM} Optical Monitor to Study Cluster Galaxy Evolution}

\author{Neal A. Miller\altaffilmark{1}} 
\email{nmiller@astro.umd.edu}

\author{Richard O'Steen\altaffilmark{2}}

\author{Steffi Yen\altaffilmark{1}}

\author{K.D. Kuntz\altaffilmark{3,4}}

\author{Derek Hammer\altaffilmark{3,4}}

\altaffiltext{1}{Department of Astronomy, University of Maryland, College Park, MD 20742}
\altaffiltext{2}{Department of Physics and Astronomy, University of North Carolina, Chapel Hill, NC 27599}
\altaffiltext{3}{Department of Physics \& Astronomy, Johns Hopkins University, Baltimore, MD 21218}
\altaffiltext{4}{Laboratory for X-ray Astrophysics, NASA Goddard Space Flight Center, Code 662.0, Greenbelt, MD 20771}

\begin{abstract} 

We explore the application of {\it XMM-Newton} Optical Monitor (XMM-OM) ultraviolet (UV) data to study galaxy evolution. Our sample is constructed as the intersection of all Abell clusters with $z < 0.05$ and having archival XMM-OM data in either the $UVM2$ or $UVW1$ filters, plus optical and UV photometry from the Sloan Digital Sky Survey and {\it GALEX}, respectively. The eleven resulting clusters include 726 galaxies with measured redshifts, 520 of which have redshifts placing them within their parent Abell clusters. We develop procedures for manipulating the XMM-OM images and measuring galaxy photometry from them, and confirm our results via comparison with published catalogs. Color magnitude diagrams (CMDs) constructed using the XMM-OM data along with SDSS optical data show promise for evolutionary studies, with good separation between red and blue sequences and real variation in the width of the red sequence that is likely indicative of differences in star formation history. This is particularly true for $UVW1$ data, as the relative abundance of data collected using this filter and its depth make it an attractive choice. Available tools that use stellar synthesis libraries to fit the UV and optical photometric data may also be used, thereby better describing star formation history within the past Gyr and providing estimates of total stellar mass that include contributions from young stars. Finally, color-color diagrams that include XMM-OM UV data appear useful to the photometric identification of both extragalactic and stellar sources.

\end{abstract}

\section{Introduction}\label{sec-intro}

The power of ultraviolet (UV) data has been demonstrated by numerous publications driven by data collected with the {\it Galaxy Evolution Explorer} satellite \citep[GALEX,][]{martin2005}. UV emission comes primarily from young stars, with the youngest and most massive stars dominating the far-UV while stars with lifetimes of about a Gyr dominate the near-UV. The addition of GALEX UV photometry to optical photometry from the Sloan Digital Sky Survey \citep[SDSS;][]{york2000} thus allows for improved classification of millions of stars and galaxies \citep{bianchi2007}. In fact, GALEX UV photometry is invaluable for identification of hot stars and white dwarfs within the Milky Way \citep{bianchi2011}. It is also vital in estimating the dust attenuation in galaxies and thus properly evaluating star formation rates \citep{treyer2007,johnson2007,salim2007}.

One of the simple and yet powerful tools in applying UV data to the understanding of galaxies and their evolution is the color-magnitude diagram (CMD). Galaxies are known to separate well into red and blue populations \citep[e.g.,][]{baldry2004}, forming a tight red sequence and a broader but still well-described blue cloud. Colors formed using near-UV data paired with optical data produce a larger separation between red and blue galaxies, thereby producing cleaner assignments of galaxies \citep{wyder2007}. More significantly, CMDs constructed using near-UV and optical data reveal a population of objects intermediate to the red sequence and blue cloud and dubbed the ``green valley'' \citep{wyder2007}. These may represent an important transitory population consisting of star-forming galaxies and active galactic nuclei (AGN) fading into quiescence and an eventual home as ``dead and red'' galaxies on the red sequence \citep{schiminovich2007,martin2007}. Alternatively, they may be rejuvenated red sequence galaxies to which a ``frosting'' of younger stars has been added \citep{trager2000}. Along these same lines, the spread in the UV-optical color at a given absolute magnitude for early-type galaxies can be explained by a significant fraction ($\sim30\%$) having experienced star formation within the past Gyr \citep{kaviraj2007}.

The {\it XMM-Newton} satellite \citep{jansen2001} is also equipped with an Optical Monitor telescope \citep[][XMM-OM]{mason2001}. This instrument provides concurrent observation of XMM's X-ray targets in a selection from among six filters, three of which approximate the Johnson $U~B~V$ filters and three of which reside in the UV ($UVW2$, $UVM2$, and $UVW1$). Although of smaller aperture and with a smaller field of view than GALEX, XMM-OM provides higher resolution images with a point spread function that is roughly a factor of three finer than that of GALEX. The $UVM2$ filter has an effective wavelength and width comparable to GALEX $NUV$, and thus can directly emulate the science performed with that filter. The XMM-OM data might also prove useful in evaluating the age and abundance of old stellar populations and hence elliptical galaxies. The region of the spectrum of a stellar population between about 2000$\mbox{\AA}$ and 3200$\mbox{\AA}$, and hence the $UVW1$ filter, should be dominated by main sequence stars near the main sequence turn-off. Photometry in this wavelength range can help break the age-metallicity degeneracy that plagues optical filter measurements by placing a strong lower limit on the metallicity of that population \citep{dorman2003}.

In this paper, we investigate the usage of XMM-OM data to study galaxies. We ultimately strive to apply the XMM-OM photometry to understand the star formation histories of cluster galaxies, and evaluate such information in relation to evolutionary processes believed to be at work in cluster environments. XMM is a popular choice for X-ray observations of galaxy clusters on account of its sensitivity to the extended emission of the hot intracluster medium from thermal bremsstrahlung, implying that there is a large and nearly untapped database of associated UV observations of galaxy clusters made by XMM-OM. Prior publications capitalizing on XMM-OM data associated with observations of galaxy clusters have focused on solely the brightest cluster galaxies and used the UV data to evaluate star formation in such objects and cluster cooling flows \citep{mittaz2001,hicks2005,donahue2010}. Our work utilizes the full field of view of XMM-OM in order to compile a database with hundreds of galaxies having UV photometry to pair with SDSS data. Unless otherwise noted, all magnitudes are in the AB system and we assume a standard $\Lambda$CDM cosmology with $\Omega_m = 0.3$, $\Omega_\Lambda = 0.7$, and $H_0 = 70$ km~sec$^{-1}$~Mpc$^{-1}$.

We describe the sample construction and data sources in Section \ref{sec-data}. Our procedures for processing the images and measuring galaxy photometry are described in Section \ref{sec-proc}, followed by the construction of multiwavelength catalogs in Section \ref{sec-catalog} and their usage as checks of calibration in Section \ref{sec-calcheck}. The remainder of the paper is devoted to applications of the XMM-OM UV data, including construction and assessment of CMDs (Section \ref{sec-cmd}), evaluation of star formation history (Section \ref{sec-sfh}), inclusion of XMM-OM UV data in photometric redshift and galaxy stellar mass estimation (Section \ref{sec-pzandm}), and selection of unusual objects via color-color diagrams including XMM-OM UV data (Section \ref{sec-nonclus}). A summary is provided in Section \ref{sec-conclude}.

\section{Sample Construction and Data Sources}\label{sec-data}

\subsection{Sample Construction}

Our goal is to produce well-calibrated UV data that can be added to optical photometry to yield more robust star formation histories of cluster galaxies. For this reason we selected clusters with coverage in both the SDSS and GALEX surveys, allowing us to piggy-back off of their calibration. We are thus able to test and calibrate the XMM-OM filters since they lie at wavelengths between the GALEX and SDSS surveys. We also imposed a cut-off in cluster systemic redshift of $z \leq 0.05$, guided by the expectation that this would produce reasonable numbers of UV-detected galaxies per cluster. A large elliptical galaxy ($M_r = -22$) at this maximum redshift would have $r \approx 14.7$ and $NUV \approx 20.7$, just fainter than the magnitude limit of the GALEX All-Sky Imaging Survey \citep[AIS, with $m_{NUV} \sim 20.5$][]{martin2005}. We correlated the XMM ``Master'' catalog, obtained through NASA's High Energy Astrophysics Science Archive Research Center (HEASARC), with the Abell catalog \citep{abell1989} using a matching radius of 8\arcmin{} to insure that the cluster cores would be represented in the 17\arcmin{} square FOV of the XMM-OM. This produced a sample of 11 clusters with either {\em UVM2} or {\em UVW1} data, $z \leq 0.05$, and coverage in both the SDSS and GALEX surveys. Table \ref{tbl-sample} provides the details of the clusters and their XMM observations.

For some of the clusters -- most notably A1656, or Coma -- XMM-OM observations of regions outside the cluster cores are also available. As these provide piecemeal coverage of the clusters and are often non-contiguous with the XMM-OM observations of the cluster cores, we have not included such data in the present study. Thus, the galaxies in our sample are exclusively those in or projected upon the cluster {\it cores}. For reference, the farthest cluster in our sample (A119) has $z_{sys} = 0.0442$ and thus 1\arcsec{} equals about 0.9 kpc while for the nearest (A1656) $z_{sys} = 0.0231$ and 1\arcsec{} is about 0.5 kpc. These translate to coverage from the core out to 0.63 Mpc (A119, diagonal of 17\arcmin{} $\times$ 17\arcmin{} XMM-OM image) at best, and 0.24 Mpc (A1656, center to edge of XMM-OM image) at worst.

\subsection{{\it XMM-Newton} Optical Monitor}

XMM-OM is a 30cm diameter Ritchey-Chr\'etien telescope that is coaligned with the X-ray telescopes of {\it XMM-Newton} \citep{mason2001}. Its filter wheel provides six broad-band imaging filters in addition to a blocked aperture, a pair of grisms, a magnifier, and a ``white'' filter. The detector is a microchannel plate intensified CCD, wherein incident photons strike a photocathode and the resulting electrons are amplified through a pair of microchannel plates. The final electrons illuminate a phosphor screen where they are registered by the $256 \times 256$ pixel CCD. The CCD is read out rapidly, once every 11 $\mu$s, and the centroids of each event are recorded to 1/8th of a pixel. The FOV is a 17\arcmin{} square, although limitations to onboard memory and centroiding require observations of this full area to be built up from a sequence of ``science windows.'' In the standard imaging mode, a small region (5\arcmin{} $\times$ 5\arcmin) at the center of the FOV is observed continuously at ``high resolution'' (i.e., keeping the centroid information to 1/8th of a CCD pixel and providing an output image with $\sim$0\farcs5 pixels) while a sequence of five science windows cover nearly the full FOV through a ``low resolution'' mosaic (using $2 \times 2$ binning and producing $\sim$1\arcsec{} pixels). Note that the PSF of an unresolved source (roughly 2\arcsec{} for the UV filters) is the same for either the high resolution or low resolution modes, with the only difference being the sampling. The mosaic pattern for this standard low resolution imaging mode is a central square surrounded by four overlapping rectangular regions \citep[see Figure 2 of ][]{mason2001}. Since these regions overlap, the integration time across the full FOV is not constant even if that in each science window is identical. In another commonly-used imaging mode, the small high resolution window is dropped and the full 17\arcmin{} field is observed through four rectangular science windows with the same output pixel size as the aforementioned low resolution mosaic \citep[``ENG-4''; see Figure 2 of ][]{kuntz2008}. For our work we use the full 17\arcmin{} ``low resolution'' images, from either the standard five science window mosaic or the four science window ``ENG-4'' mode, to encompass greater areas and consequently obtain photometry on larger numbers of objects.

As noted previously, three of the XMM-OM broad-band filters have effective wavelengths placing them within the UV portion of the spectrum (see Figure \ref{fig-filts}). The shortest wavelength filter, {\em UVW2}, has poor throughput, suffers from a red leak \citep{talavera2008}, and is not substantially different from the {\em UVM2} filter in terms of effective wavelength and width. Consequently, in this work we focus our efforts on data collected using the {\em UVM2} and {\em UVW1} filters.

\subsection{SDSS}

The Sloan Digital Sky Survey \citep[SDSS;][]{york2000} is an imaging and spectroscopic survey mainly focused on the northern Galactic cap. The 2.5m diameter telescope used for the survey has a wide ($3^\circ$) field of view and a camera that images in five broad-band filters ({\em u}, {\em g}, {\em r}, {\em i}, and {\em z}) simultaneously through drift-scan observations \citep{gunn1998}. The images typically have 1\farcs5 resolution and have a 95\% completeness limit at about $r = 22.2$ \citep{abazajian2004}. Spectra are then collected based on the imaging catalogs, using color information to select targets likely to belong to a range of source classes (e.g., normal galaxies, luminous red galaxies, quasars, etc.). The SDSS spectrograph uses drilled plug plates with 3\arcsec{} fibers to collect 640 spectra in a single field. The SDSS has been released to the public through a series of data releases; a comprehensive description of the data can be found in the Early Data Release \citep{stoughton2002} with revisions to the determination of key parameters noted in subsequent releases up through the recent Seventh Data Release \citep[DR7;][]{abazajian2009}. Each data release is cumulative, and we use the DR7 for the present work.

\subsection{GALEX}

GALEX is a 50cm diameter Ritchey-Chr\'etien telescope dedicated to imaging and grism spectroscopy in the ultraviolet \citep{martin2005}. Like XMM-OM, it also uses microchannel plates and photocathodes for source detection although separate systems for {\em NUV} and {\em FUV} allow it to image both filters simultaneously (see Figure \ref{fig-filts} for filter curves). The GALEX resolution is more coarse than that of XMM-OM \citep[4\farcs5 - 6\arcsec ;][]{morrissey2005} on account of several factors including both the instrument itself and how the data analysis pipeline reconstructs photon events. The large effective area of GALEX and its much larger field of view (1\fdg25) significantly differentiate it from XMM-OM.

The GALEX data pipeline faces the complicated task of translating photon positions and pulse heights into a corrected output image upon which source detection and photometry may be performed; \citet{morrissey2007} provides a thorough description. Data are collected through a 1\farcm5 spiral dither pattern in order to smooth over variations in the flat field and avoid having bright sources striking the same regions of the microchannel plates for any length of time. Observations of white dwarf standards have shown that GALEX magnitudes are accurate to an rms of about 0.05 in {\em FUV} and 0.03 in {\em NUV}. Once the final flux calibrated and astrometrically registered image has been created, catalogs are constructed using the Source Extractor \citep[``SExtractor,''][]{sextractor} package. Our data correspond to the fifth GALEX data release (GR5).

In addition to guest investigator observations, GALEX performed several surveys \citep{martin2005,morrissey2007}. The AIS, mentioned previously, is intended to cover the entire sky and has limiting magnitudes of $m_{FUV} \sim 19.9$ and $m_{NUV} \sim 20.8$ through short exposures (100 seconds). The Medium Imaging Survey (MIS) increases the integration time to about 1500 seconds and depth to $m_{FUV} \sim 22.6$ and $m_{NUV} \sim 22.7$. It will cover the roughly one-quarter of the sky coincident with the SDSS. There are also a Deep Imaging Survey (DIS) and an Ultra-Deep Imaging Survey (UDIS) for several fields of interest determined through existing multiwavelength surveys, a Nearby Galaxies Survey (NGS), and multiple spectroscopic surveys typically centered on DIS or UDIS regions.

\section{XMM-OM Image Processing and Photometry}\label{sec-proc}

\subsection{XMM-OM Image Processing}

HEASARC contains the low resolution images as processed by \citet{kuntz2008} for the construction of OMCat. These were produced using SAS\footnote{Version 6.5.0. SAS was developed by members of the {\it XMM-Newton} Science Survey Centre, a consortium of 10 institutions led by Prof. M. Watson of the University of Leicester.}, and the individual science windows have been stitched together to produce the full $\sim$17\arcmin{} square images. They have also been corrected to match the astrometry of the USNO-B1 catalog \citep{monet2003}. We obtained these images and used them as the starting point for our processing. Table \ref{tbl-sample} includes the XMM ObsID's and other summary information. 

The source detection and photometry procedures of SAS are designed for point sources and not extended objects like galaxies, and thus we do not use OMCat photometry nor do we use SAS for such purposes. This produces some extra complications in our image processing, as SAS automatically performs some corrections when performing photometry. Most notable among these are corrections for coincidence loss and time sensitivity degradation. The former is a reflection of the photon counting nature of the system, as multiple photons may arrive at the detector within the brief time between reads of the CCD. Fortunately, this is only a problem for very bright and unresolved sources and amounts to about a 10\% effect for stars with $m_{UVW1} \approx 16.1$ and $m_{UVW1} \approx 14.7$ in Vega magnitudes \citep[$m_{UVM2} \approx 14.9$ and $m_{UVM2} \approx 13.3$ in AB magnitudes; the OM Calibration Manual][]{talavera2008}. These magnitudes are roughly the same as the integrated magnitudes of the brightest galaxies in our sample, which have spread their flux over tens of arcseconds. We therefore make no explicit corrections for coincidence loss (see Section \ref{sec-calcheck} for a discussion on the validity of this assumption). Sensitivity degradation is the decline in throughput of XMM-OM over time presumably through degradation of its photocathode. Regular monitoring of a collection of standard stars has shown that this degradation can be fit linearly such that the correction factor is equal to $A + B \times MJD$, where $MJD$ is Modified Julian Date. The slope and intercept pairs are determined for each filter and provided in \citet{talavera2008}\footnote{More recent monitoring has shown that post-2008 the loss of sensitivity over time has slowed. The most recent observational data we use were collected in 2006, so we adopt the original correction factors.}. For our data, the correction factors range from about 1.07 to 1.13 for the {\em UVM2} observations and from about 0.98 to 1.07 for the {\em UVW1} observations. Thus, the total counts for each observation can be corrected to a standard ``zero epoch'' for accurate photometry. 

The SAS-produced HEASARC images are normalized to units of count rate per 1000 seconds, regardless of actual integration time. We split off the image extension corresponding to the pixel-by-pixel integration time, multiplied this exposure map with the science image, divided by 1000 to convert the units to net counts, and multiplied by the correction factor to account for time sensitivity degradation. In several cases, there were multiple XMM-OM observations in a single filter for a given galaxy cluster. Having corrected the count rate of each to the same zero epoch, we simply translated them to the same astrometric grid and summed their counts. For A1656, we masked out the large diffraction spikes and artifacts associated with a bright star prior to performing the sum. As the roll angle of the observation was different for each A1656 observation, this masking does not remove sky coverage in the final image. The same translation to a common astrometric grid and sum (and masking for A1656) was performed on the exposure maps, and our science images are the net count maps divided by the net exposure maps (hereafter we will refer to this image, with units of counts per second, as our ``science image''). The final step was additional astrometric correction, as described below in Section \ref{sec-astrom}.

\subsection{Source Detection and Photometry}

As with GALEX data, SExtractor was used to perform source detection and photometry. SExtractor was run in the ``pseudo-dual-image'' mode where both the detection and measurement images were the science image, with the advantage being that this mode allows the use of separate weightings for detection and measurement. In this way we are able to detect sources using the science image but compute their errors properly based on the exposure map. Thus, a single zero point magnitude appropriate for an image with units of counts per second can be applied yet regions with deeper total exposure will have smaller errors. This is accomplished by setting the ``GAIN'' parameter to the average exposure time (in units of seconds) of the image. For the zero points, we used 17.41 for the {\em UVM2} filter and 18.57 for the {\em UVW1} filter as specified in \citet{talavera2008}. The background map, subtracted from the image for performing photometry, was determined locally from the science image itself using the standard SExtractor estimation which uses the mean or an estimate of the mode based on the mean and median, depending on the background distribution. For images with very low background rates this is erroneous as the background is better described as a Poisson distribution, as done in the GALEX pipeline \citep{morrissey2007}. Our typical background levels ($\approx 2 \times 10^{-3}$ cts~s$^{-1}$~arcsec$^{-2}$ for {\em UVM2} and $\approx 3 \times 10^{-3}$ cts~s$^{-1}$~arcsec$^{-2}$ for {\em UVW1}) are slightly higher than those in GALEX data because the XMM-OM filters are at longer wavelengths, and our comparison of photometry suggests any resulting magnitude errors are small (see Section \ref{sec-calcheck}) so we relied upon the standard SExtractor estimation. We did adjust the background mesh size to 48 pixels and use $3 \times 3$ pixel median filtering, similar to the parameters used in the GALEX pipeline after accounting for the differences in instrumental resolution. The background mesh size was chosen through several trials, and is large enough not to be overly influenced by extended sources yet fine enough to handle real variations (for example, regions of deeper total integration caused by the mosaic pattern of the default XMM-OM image acquisition or stacking of multiple images). We also found it valuable to create and use a flag image, since the initial mosaic images are rectangular with the standard North-up, East-left orientation with the measured pixels covering only a subset of the full area. The input flag image thus had all non-covered pixels, plus those within about 4\arcsec{} of a non-covered pixel, set to one while pixels covered by the XMM-OM observation were set to zero.

Our main output is the ``MAG\_AUTO,'' or Kron-like, aperture magnitudes along with simple source characterization (position in both pixels and RA/Dec, Kron radius, star/galaxy classification, and flags). The choice of MAG\_AUTO over aperture magnitudes was driven mainly by the GALEX pipeline also producing MAG\_AUTO magnitudes, and that these are the standard ones applied for deriving colors in combination with SDSS data \citep[e.g.,][]{bianchi2007,wyder2007}. The star/galaxy classification assumed that the PSF of the {\em UVM2} images was 1\farcs8 while that of the {\em UVW1} images was 2\arcsec{} \citep{talavera2008}, although the XMM-OM-based star/galaxy classification was generally ignored as we associate our sources with SDSS counterparts and rely on the SDSS photometry for identification of stars (see Section \ref{sec-catalog}). 


\subsection{Astrometric Correction}\label{sec-astrom}

Even when an external catalog such as the USNO-B1 is used, the astrometric correction within SAS can often fail \citep{kuntz2008}. Furthermore, our intention to combine our UV catalogs with exisiting optical and UV data made it a natural choice to tie our astrometry to the SDSS. We correlated our initial XMM-OM catalogs with the SDSS catalog, separately for {\em UVM2} and {\em UVW1}, assigning each XMM-OM source a nearest neighbor from the SDSS. We then selected all XMM-OM/SDSS matches with positional separations less than 3\arcsec, SDSS type designations as stars, and no flags in their XMM-OM photometry. These objects were then used to correct the astrometry of the {\em UVM2} and {\em UVW1} images to the same frame as the SDSS. The fit allowed shifting of both the $x$ and $y$ coordinates (while maintaining the assumption that the axes remained perpendicular), as well as field rotation. The procedure was iterative in that we performed the transformation, inspected the residuals for the individual stars that were used in it, and removed those with large residuals. Many of these objects with high residuals are saturated stars with poorly-identified centers or mis-matched objects. For the typical {\em UVM2} images, the final fits were based on about 25 stars spread over the full field and not including the 5 to 10 that were removed, while for the {\em UVW1} images these numbers were on the order of 100 and 15 to 20. Shifts of about an arcsecond were typical, with all rotations very close to zero ($\lesssim0.05$ degrees). For one of the A2199 images (XMM ObsID 0008030301, {\em UVW1} filter) the initial astrometry was off by more than 3\arcsec{} and a preliminary round of astrometric correction was required. Figure \ref{fig-astrom} depicts the resulting improvement in separation between XMM-OM objects and their nearest SDSS counterpart (both stars and galaxies). Once the astrometric correction was determined, it was applied to each image for that cluster and filter to produce final images (i.e., the science image with units of counts per second, the net exposure map, and the flag image). All source detection and photometry procedures were then repeated to produce the final XMM-OM object catalogs.

\section{Multiwavelength Catalogs and Calibration}

\subsection{Construction of Multiwavelength Catalogs}\label{sec-catalog}

Based on the results just shown in Figure \ref{fig-astrom}, we accepted as matches all XMM-OM catalog objects that had an SDSS counterpart within 2\arcsec . We then adopted the SDSS position to do all further associations. In some cases, multiple SDSS objects were associated with a single XMM-OM object in which case we manually inspected each association to select the most likely counterpart (usually the brightest SDSS object). Once an association had been made, the SDSS model magnitudes \citep[model magnitudes are recommended for determining colors of extended objects;][]{abazajian2004} and errors were saved along with the SDSS position and type designation (i.e., star or galaxy). After the final matched XMM-OM/SDSS catalog was constructed, we searched for GALEX counterparts within 4\arcsec{} of each object's SDSS position \citep[e.g., see ][ for justification of this matching radius]{bianchi2007}. When multiple GALEX tiles covered the position, we selected the measurements corresponding to the tile with the longer integration time. As with the XMM-OM SExtractor photometry, we used the GALEX ``MAG\_AUTO'' {\em FUV} and {\em NUV} magnitudes and errors. The pairing of SDSS model magnitudes and GALEX MAG\_AUTO ones is consistent with the majority of work that investigates UV/optical colors of galaxies \citep[e.g.,][]{bianchi2007,wyder2007}. Similarly, the \citet{blanton2007} \kcorrect software which we will use in later analysis relies on these measurements. 

Finally, we collected publicly-available redshifts for the matched sources using the SDSS as our primary spectroscopic data source and the NASA/IPAC Extragalactic Database (NED) as our secondary source. That is, if an object had both an SDSS redshift and one from NED we adopted the SDSS value. In the case of the NED associations, an allowable matching radius of 6\arcsec{} was used. We further augmented our redshift list by using the Marzke et al. (in preparation) spectroscopic catalog for Abell 1656, which is also based on SDSS object positions. In addition to the redshift, we save its source (SDSS, NED, or other) and SDSS identifier (Plate, MJD, and Fiber) where applicable.

In summary, our data catalog for analysis includes photometry from nine filters and their errors: {\em FUV, NUV, UVM2, UVW1, u, g, r, i,} and {\em z}. It is XMM-OM selected, in that the prerequisite for inclusion is the presence of either a {\em UVM2} or {\em UVW1} measured magnitude and a matched entry from the SDSS. We do include objects with flags on their XMM-OM photometry, as these flags often indicate benign considerations such as noting that an object was deblended or relatively close to an image boundary. The coordinates for each object in the catalog are those from the SDSS, although we do maintain the measured XMM-OM positions for comparison. Finally, any spectroscopically-measured redshift for each object is also included in its catalog entry.

Table \ref{tbl-nums} summarizes the multiwavelength catalog. In total, we have 3,311 XMM-OM-detected objects with counterparts in the SDSS. Most individual clusters have around 300 objects, with A119 and A400 having signficantly fewer (64 and 58, respectively). In the case of A119, this is partially a reflection of its Southern declination as the SDSS only covers about half of the area in the XMM-OM images. In addition, the {\em UVW1} integration for A119 is the shortest of all clusters with such data. As this filter has greater transmission than {\em UVM2} and coincides with an intrinsically brighter portion of the SED of most galaxies, there are typically three or four times as many {\em UVW1} detections per cluster as there are {\em UVM2} detections for equal integration times. This is also the explanation for the lesser number of detected objects in A400, as this cluster has only {\em UVM2} data and no coverage with {\em UVW1}. A1656 has the largest number of objects (652) on account of its deep coverage (i.e., the co-addition of seven or eight separate XMM-OM observations) in both {\em UVM2} and {\em UVW1}.

Approximately one-quarter of all XMM-OM-detected objects in our sample have an available redshift (768/3311, or 23\%). Of these, the majority (726) are galaxies with $0.001 < z < 0.999$ which will be used in SED fitting and associated analysis (Section \ref{sec-extinct} and Section \ref{sec-apply}). Most of the 42 objects with redshifts outside of this range are stars, although up to six possible quasars are detected: three certain quasars with SDSS spectra, two objects with redshifts from NED, and one SDSS object with an ``uncertain'' spectral classification. We will discuss these objects further in Section \ref{sec-apply}. For simplicity, we isolate the potential cluster members by applying cuts of $z_{sys} \pm0.01$ as this is approximately 3$\sigma$ for a rich cluster having a velocity dispersion of 1,000 \kms . This yields 518 cluster galaxies in the catalog. For A2197, the lesser number of galaxies with measured redshift (and measured redshift placing them within the cluster) is likely due to the XMM-OM observation being offset from the cluster core by over 6\arcmin .

\subsection{Comparison with Other Catalogs}\label{sec-calcheck}

With resources such as OMCat and GALEX available, we take advantage of their catalogs to provide checks on our photometry and image processing procedures. First, we match our photometry catalogs with those of OMCat as shown in Figure \ref{fig-omcatck}. All matches with SDSS type classification of star and an OMCat counterpart within 2\arcsec{} are plotted, with our photometry modified to reflect that the OMCat photometry is in the Vega system \citep[we used $m_{AB} - m_{Vega}$ of 1.64 and 1.37 for {\em UVM2} and {\em UVW1}, respectively, as taken from ][]{talavera2008}. For clusters with multiple XMM-OM observations in a given filter, the OMCat photometry is based on the individual observations whereas our photometry has been performed on images that have combined all the data. Figure \ref{fig-omcatck} does ``double count'' galaxies in this regard, as for such systems we have used the OMCat data from each one of these observations. It can be seen that the overall consistency of the photometry is good, although with slight offsets in the zero points such that our measurements are less than 0.1 mag fainter. The large outliers at magnitudes brighter than about $m = 19$ for $UVW1$ are primarily drawn from two clusters, A1656 and A2052, that have eight and three XMM-OM $UVW1$ observations, respectively. In nearly all cases the discrepant points are caused by a single XMM-OM observation representing an outlier relative to the photometry of the others. This is especially apparent for A2052, where two observations are very short (800 and 900 seconds) and it is these two observations that produce the vast majority of the outliers. The third observation of A2052 has an integration of 2000 seconds and not surprisingly we find that its OMCat photometry is more consistent with our photometry.

Figure \ref{fig-omcatck} also represents a confirmation that we can safely ignore coincidence loss for our galaxy photometry. The SAS-produced photometry of the OMCat includes the correction for coincidence loss, whereas our photometry does not. We would therefore expect the relative difference between our photometry and OMCat to increase at brighter magnitudes, with our measured magnitudes being fainter. This trend is indeed seen for $m_{UVW1}^{OMCat} \lesssim 14$, while none of the {\em UVM2} comparison stars are bright enough for coincidence loss to be a factor. 

\begin{figure}
\figurenum{1}
\epsscale{0.9}
\plotone{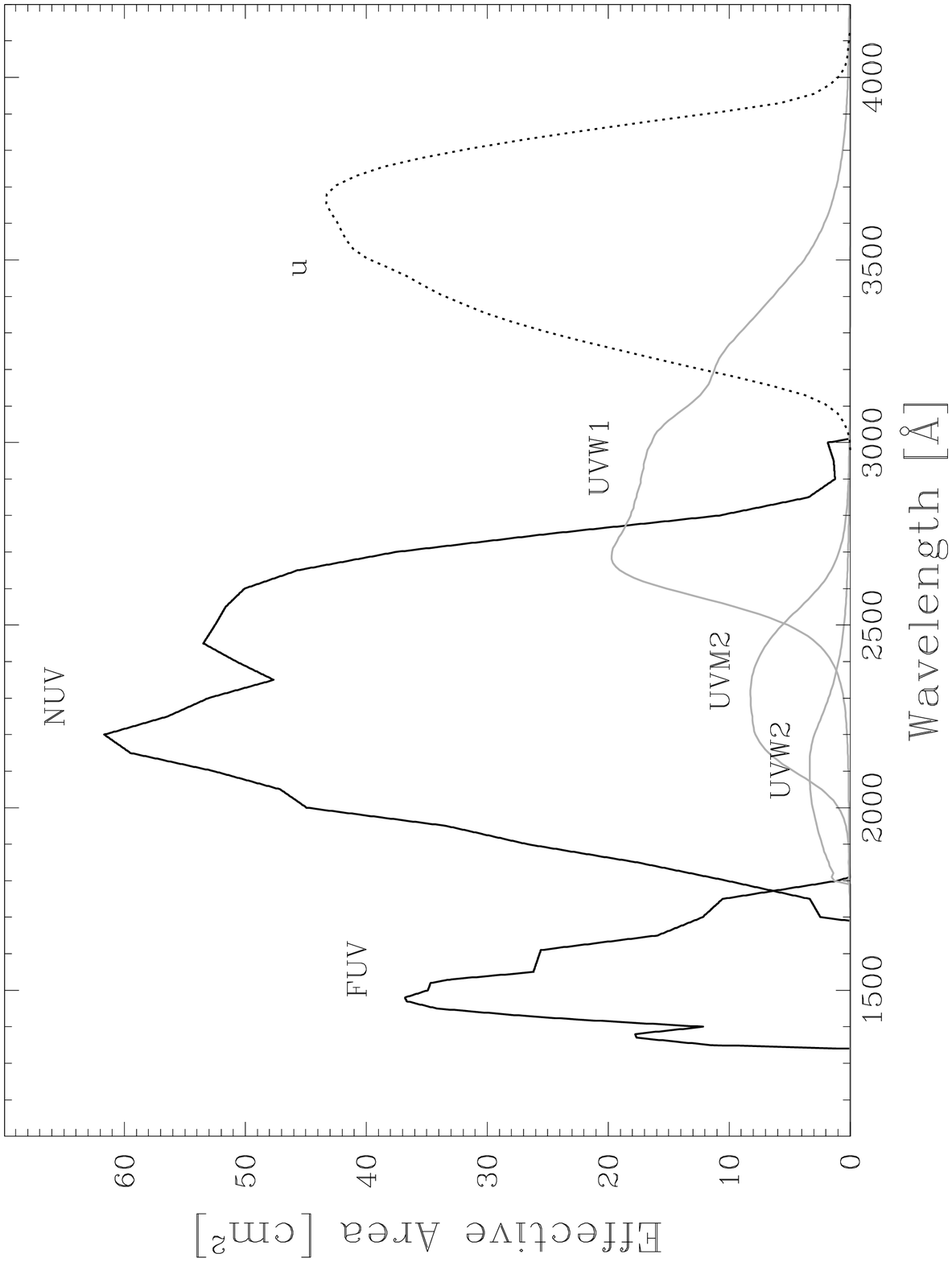}
\caption{Filter curves for XMM-OM, based on measurements from \citet{talavera2008}, with GALEX $FUV$ and $NUV$ plus SDSS $u$ also shown. The SDSS effective area assumes observation of a source at 1.3 airmass and has been divided by an arbitrary amount to fit the scale of the plot. The effective wavelengths for the filters are: GALEX $FUV$ 1528 $\mbox{\AA}$, GALEX $NUV$ 2271 $\mbox{\AA}$, $UVW2$ 2070 $\mbox{\AA}$, $UVM2$ 2314 $\mbox{\AA}$, $UVW1$ 2901 $\mbox{\AA}$, and SDSS $u$ 3551 $\mbox{\AA}$.\label{fig-filts}}
\end{figure}
\begin{figure}
\figurenum{2}
\epsscale{0.9}
\plotone{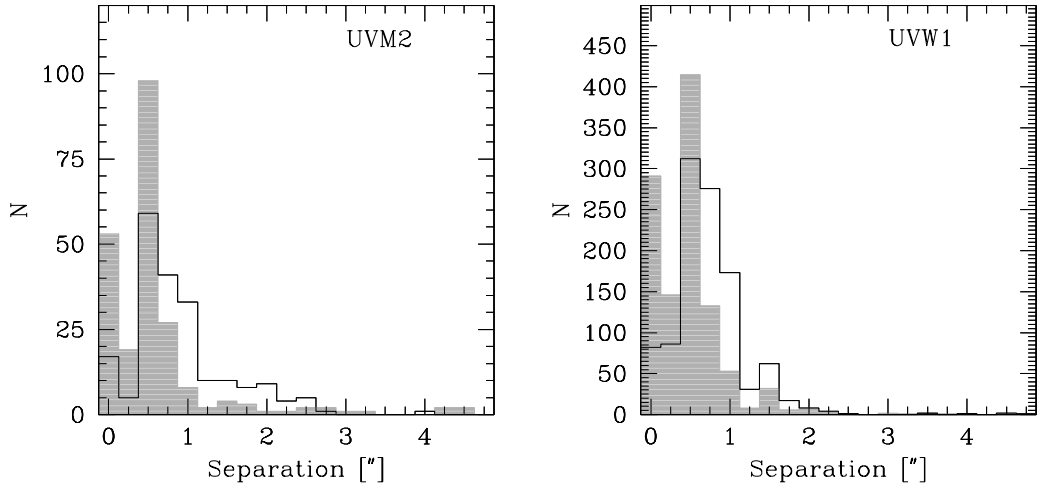}
\caption{Separations between XMM-OM object and nearest neighbor in the SDSS ($UVM2$ at left, $UVW1$ at right). The solid line and unshaded histogram represents the data prior to tying the astrometry to the SDSS, while the grey histogram represents the final astrometrically-corrected data.\label{fig-astrom}}
\end{figure}
\begin{figure}
\figurenum{3}
\epsscale{0.9}
\plotone{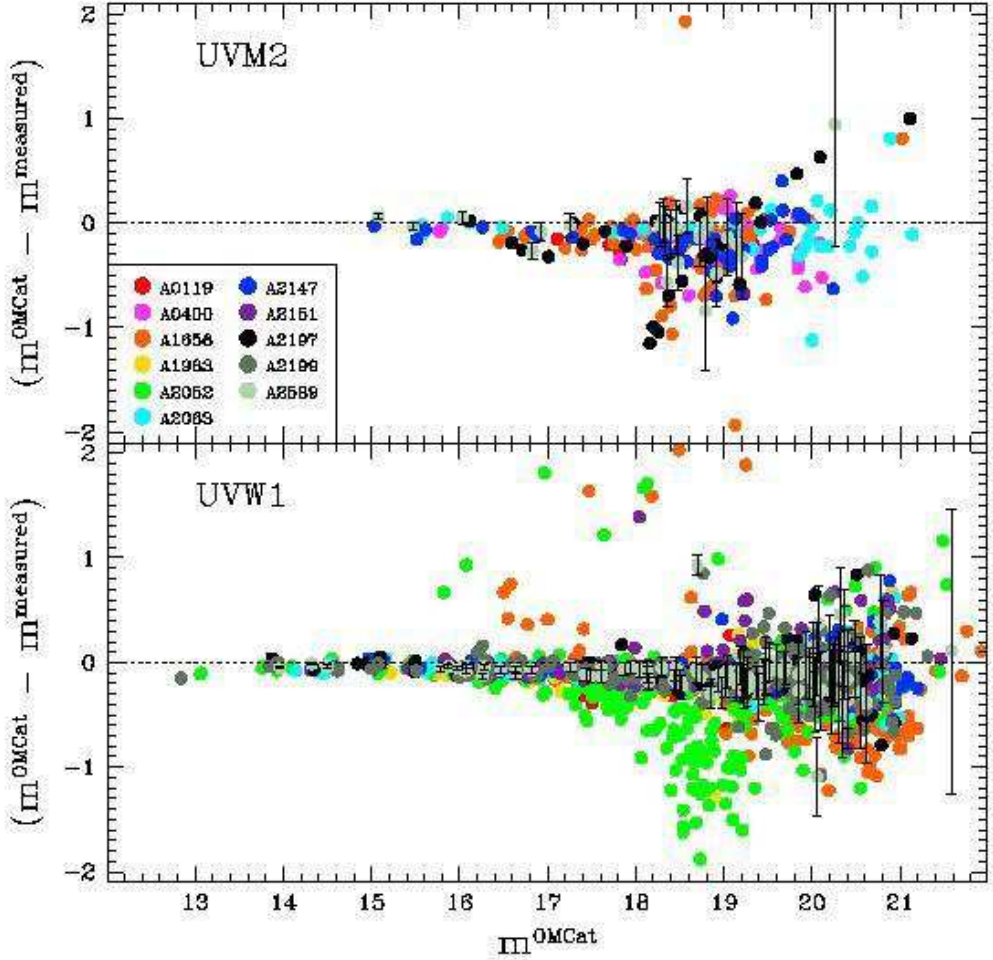}
\caption{Photometric comparison between OMCat \citep{kuntz2008} and our image processing and photometry. Magnitudes are in the Vega system, and the data points are color-coded by Abell cluster as shown in the legend (note that these same color codings will be used in all subsequent figures). Error bars are plotted for A2589 only to improve clarity, and in general are smaller than the point sizes for magnitudes $m_{UVM2} \lesssim 17$ and $m_{UVW1} \lesssim 18.5$. At fainter magnitudes, the general flaring about a consistent offset between the two catalogs is consistent with the increasing magnitude errors.\label{fig-omcatck}}
\end{figure}
\begin{figure}
\figurenum{4}
\epsscale{0.9}
\plotone{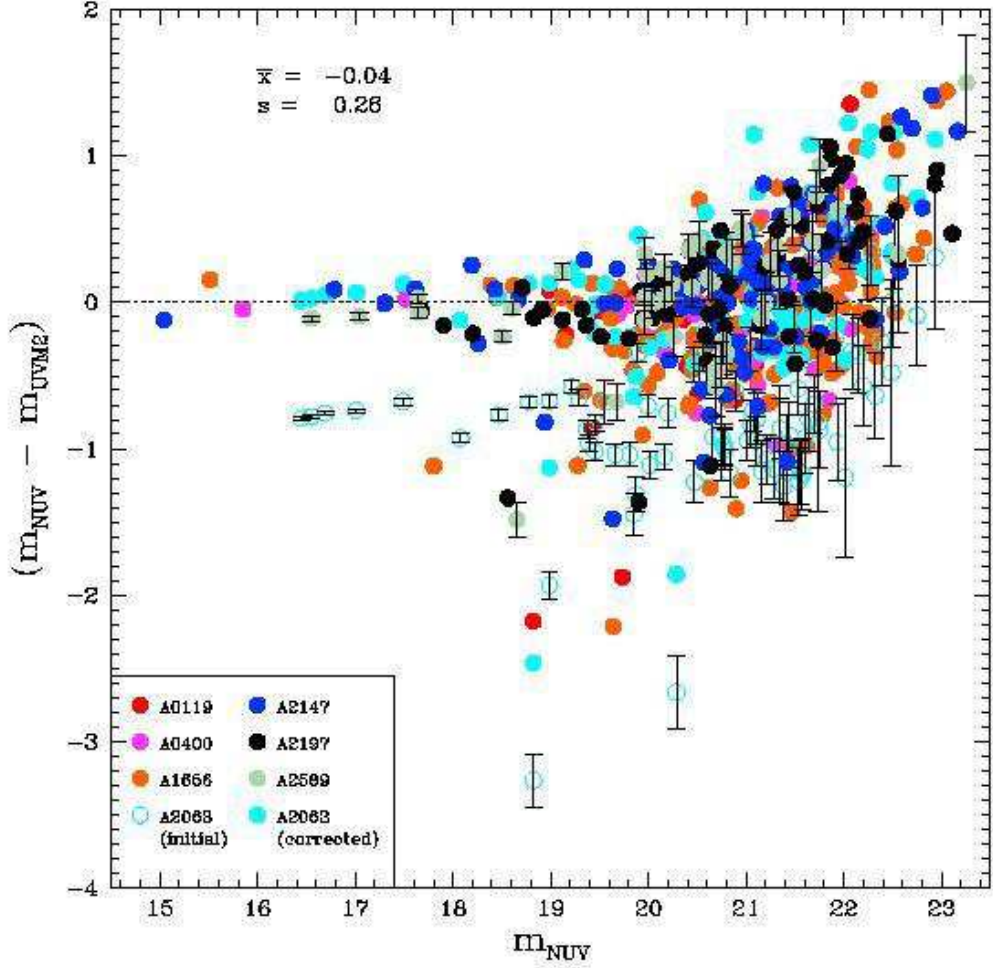}
\caption{Photometric comparison between GALEX $NUV$ and our $UVM2$ photometry (AB magnitudes). Error bars are shown for only A2063 (pre-correction) and A2589 for clarity. Note the obvious discrepancy in the photometry for A2063, which is corrected by changing the exposure time from 8348 seconds to 3976 seconds. The mean offset between the $NUV$ and $UVM2$, evaluated for sources with $m_{NUV} \leq 20$ and removing outliers, is indicated at the top left of the plot along with the dispersion in this value.\label{fig-galexck}}
\end{figure}

The GALEX {\em NUV} filter has similar effective wavelength and coverage to {\em UVM2} (refer to Figure \ref{fig-filts}) so for clusters with {\em UVM2} data we can also do a direct comparison of our photometry with GALEX. In this comparison we are not restricted to using only the photometry for stars, and both stars and galaxies are included. The result is shown in Figure \ref{fig-galexck}, where overall consistency is again demonstrated albeit with increased scatter. However, this exercise reveals that something is amiss for one cluster, A2063, that has {\em UVM2} magnitudes that appear too faint by about 0.7 mag. We have already shown that our photometric procedure produces consistent results with OMCat for the {\em UVM2} data for this observation, and in subsequent analysis we ascertain that the GALEX photometry is correct and the problem lies in the {\em UVM2} data (see Section \ref{sec-apply}). Since the offset appears to affect all of the data and not just specific objects, a simple explanation such as incorrect exposure time (too long by a factor of $\sim2$) might apply, and indeed a query to the XMM-OM calibration team confirmed that an error in the house-keeping parameter files resulted in this observation listing an exposure of 8348 seconds when in actuality it should be 3976 seconds. This results in a magnitude error of $2.5\log (8348/3976) = 0.81$, which we have included in subsequent analysis. As with the OMCat comparison, there are a handful of outliers beyond those associated with A2063. These are exclusively for objects that have {\em UVM2} magnitudes fainter than expected from their GALEX {\em NUV} magnitudes, and the majority have flagged GALEX data corresponding to detector bevel edge and window reflections. Including the corrected A2063 photometry, there are 111 objects with $m_{NUV} \leq 20$ and the mean difference between their {\em NUV} and {\em UVM2} photometry is only 0.04 mag ({\em UVM2} is fainter) but with a dispersion of 0.26 mag. Thus, the difference is not significant and our treatment of the background during source extraction appears not to have strongly biased our results for these brighter galaxies.

\begin{figure}
\figurenum{5}
\epsscale{0.9}
\plotone{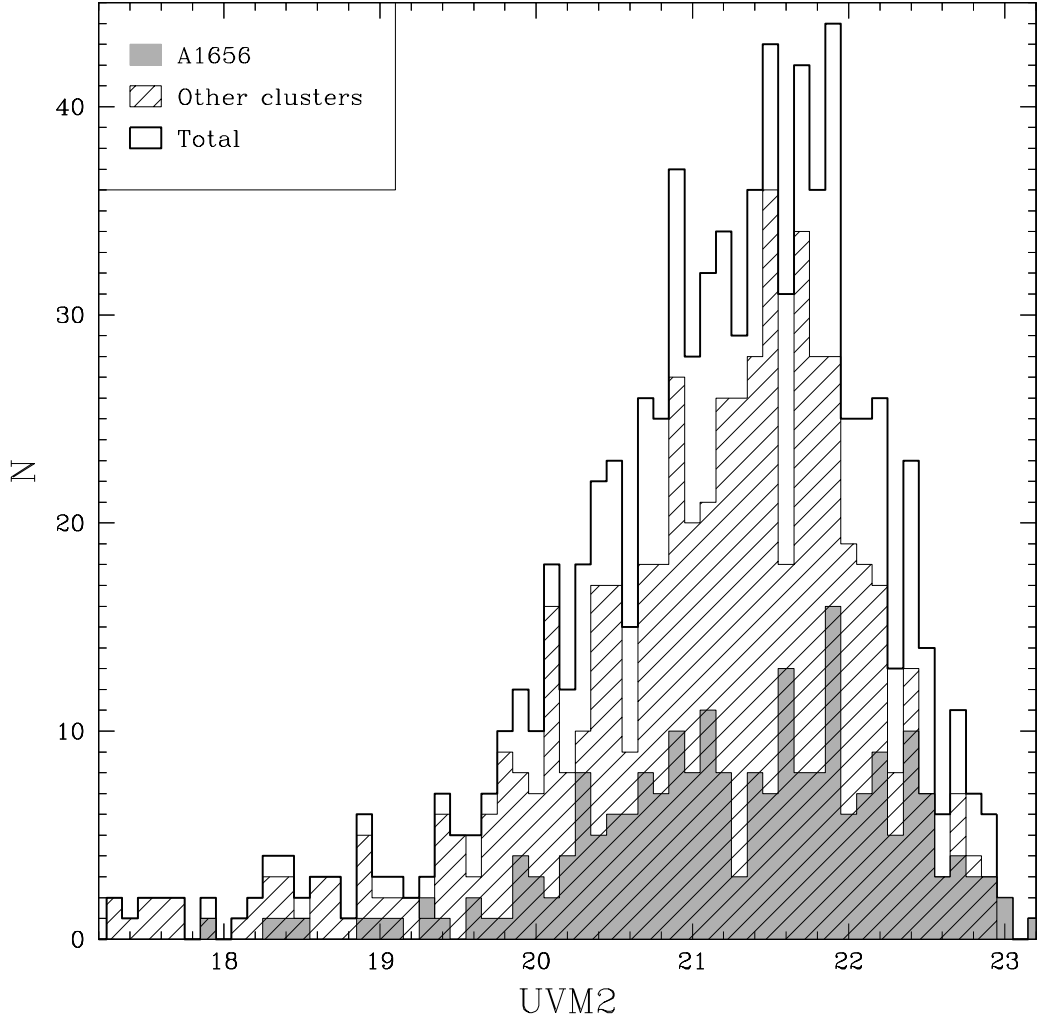}
\caption{Magnitude histogram for all objects detected in the $UVM2$ filter. The data for A1656 are indicated as a grey-shaded histogram, while the other clusters (A119, A400, A2063, A2147, A2197, and A2589; where magnitudes for A2063 have been adjusted to reflect the correct exposure time) are represented by the hatched histogram. The solid black line is the total.\label{fig-maghist}}
\end{figure}

It is also instructive to compare the depth of our {\em UVM2} data with the various GALEX surveys. In Figure \ref{fig-maghist} we plot the histogram of measured {\em UVM2} magnitudes for our cluster sample, including both unresolved and extended objects. Noting that over 30\% of the objects correspond to the A1656 data, we have separated these data in the plot and show a solid grey histogram for A1656, a hatched histogram for all of the remaining clusters, and an unshaded black histogram for the sum of the two. It can be seen that the peak in the magnitude distribution occurs somewhere fainter than $UVM2 \approx 22$. This is comparable and even slightly deeper than the GALEX All-Sky Imaging Survey (AIS), yet much more shallow than the GALEX Medium Imaging Survey (MIS) with limiting magnitudes of $m_{NUV} \sim 20.5$ and $m_{NUV} \sim 23.5$, respectively \citep{martin2005}. This roughly matches the expectations based on the effective areas of XMM-OM and GALEX (refer to Figure \ref{fig-filts}) and the respective integration times. The AIS is based on 100-second integrations, implying that XMM-OM requires several thousand seconds of integration to equal it.


\subsection{Empirical Calibration and Extinction Corrections}\label{sec-extinct}

Having produced a multiwavelength database consisting of over 700 galaxies, we can also assess how our measured {\em UVM2} and {\em UVW1} magnitudes compare with values that would be predicted based on their SEDs. For this purpose, we use the \citet{blanton2007} \kcorrect (v4.1.4) software. \kcorrect uses non-negative matrix factorization to determine the SED that best fits the photometric data for a galaxy. It is based on nearly 500 input template SEDs for instantaneous bursts of star formation, with the templates spanning a range of ages, metallicities, and dust extinctions. The input templates also include models for emission from ionized gas. This host of input templates is boiled down to five templates for the fitting of actual galaxies through matching to a set of thousands of real galaxies with GALEX and SDSS photometry. These five templates are each linear combinations of the hundreds of input templates. Thus, an actual galaxy is fit as the non-negative linear combination of five templates, which in turn have been based on hundreds of input star formation histories. \kcorrect is able to use this best-fitting SED for any given galaxy to determine K corrections in given filters, reproject onto different filter bandpasses, and determine quantities such as total stellar mass and star formation history.

Of our 726 XMM-OM selected galaxies with available redshifts, 253 have GALEX photometry for both the {\em FUV} and {\em NUV} filters to go along with their five-filter SDSS photometry. Consequently, these galaxies have SEDs that are sampled at a large range of wavelengths and to either side of the XMM-OM filters. We can therefore use \kcorrect to fit their SEDs, reproject these fits onto the XMM-OM filters, and compare the predicted magnitudes with those we have measured from the real images. This also provides a good handle on how well the fitted SEDs match the real galaxy spectra by comparing the predicted and measured magnitudes for the non-XMM-OM filters.

The magnitudes input to \kcorrect need to be corrected for Galactic extinction, so we first evaluate the reddening $E(B-V)$ for each galaxy using the \citet{schlegel1998} dust maps and convert these to extinctions using the standard $A_{X}/E(B-V) = R_X$ where $X$ represents the given filter. For the GALEX data, we then use $R_{FUV} = 8.3$ and $R_{NUV} = 8.2$ as in \citet{wyder2007} while for the SDSS we use the standard 5.155, 3.793, 2.751, 2.086, 1.479 for the {\em u}, {\em g}, {\em r}, {\em i}, and {\em z} magnitudes, respectively. These values are applicable for fairly normal galaxies with some current star formation, and since all but one of our clusters have $E(B-V) < 0.05$ any differences in extinction by galaxy type will be minor (A400 has $E(B-V) \approx 0.18$). We also follow the prescriptions of \citet{blanton2005} and perform slight corrections to the SDSS magnitudes to place them on the AB system, with $m_{AB} - m_{SDSS} = -0.036, 0.012, 0.010, 0.028, 0.040$, respectively for the {\em u~g~r~i~z} data. Finally, as recommended for \kcorrect we add small error terms in quadrature to the measured values: 0.02 for {\em FUV, NUV, g, r}, and {\em i}; 0.05 for {\em u}; and 0.03 for {\em z}.

\begin{figure}
\figurenum{6}
\epsscale{0.9}
\plotone{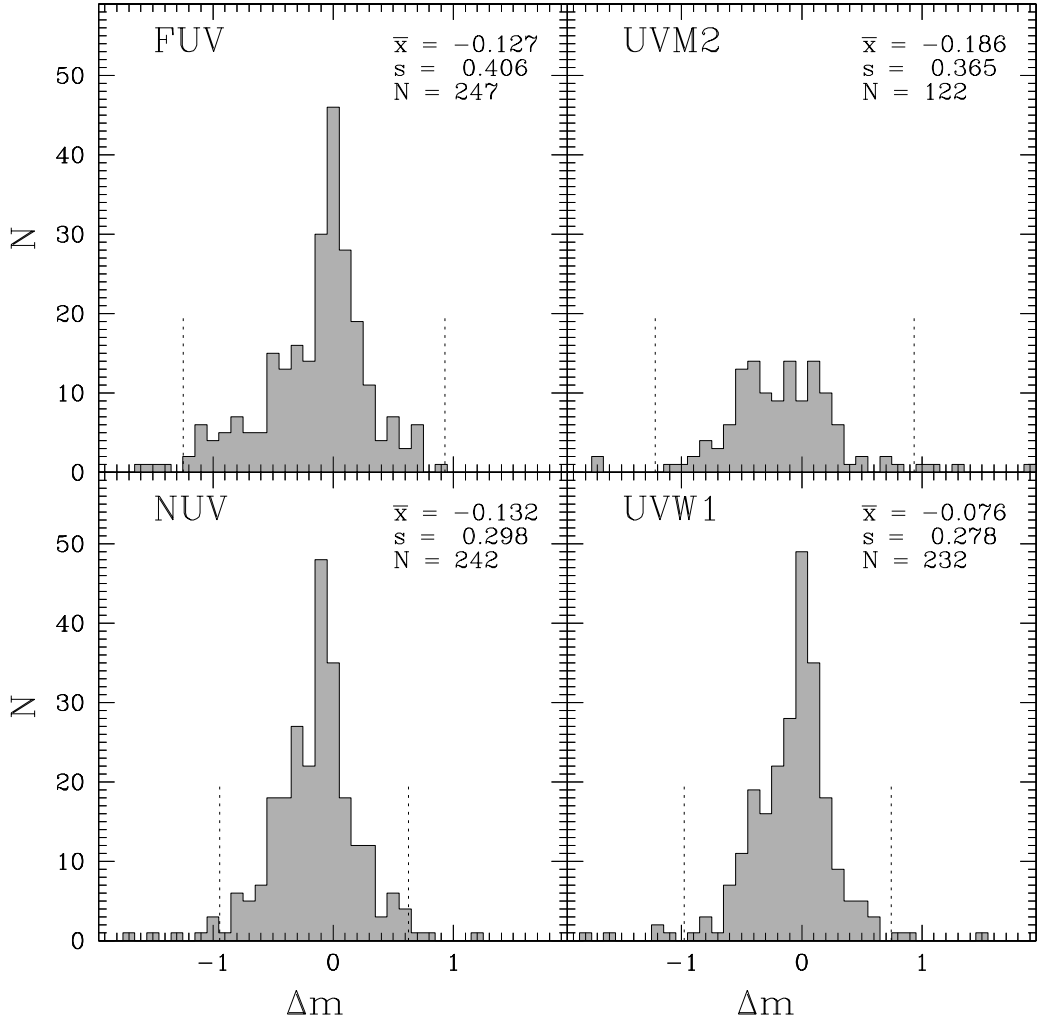}
\caption{Histograms of offsets between measured magnitudes and those predicted using \kcorrect . The mean, deviation, and number of objects are indicated at the top right of each panel and are calculated after objects differing from the main distribution by more than $3\sigma$ are removed (this range is shown as vertical dashed lines).\label{fig-extinct}}
\end{figure}

The results for the mean and dispersion of the difference between the measured and predicted magnitudes are provided in Table \ref{tbl-ecal}, and Figure \ref{fig-extinct} graphically depicts the comparison for the GALEX filters as well as the two XMM-OM filters. For each filter, we perform a single round of clipping where we remove objects with actual minus predicted magnitudes differing by more than 2$\sigma$ from the overall mean. This serves to eliminate objects that have incorrect photometry or are very poorly fit by \kcorrect . Examples of the former class of error include problems with deblending objects within the GALEX and the SDSS pipelines, while the latter can include AGN with strong non-stellar contributions to their spectra and galaxies with incorrect redshifts. It can be seen that as a whole the \kcorrect{} fits are good approximations to the actual measured magnitudes, particularly for the SDSS filters for which \kcorrect{} was designed. There do appear to be offsets for the GALEX and XMM-OM UV filters, but the sign and magnitude of these are consistent and likely related to the particulars of the SED fitting and its application to the UV portion of the spectrum, especially as a function of the morphological mix of the galaxies. We will discuss this further in Section \ref{sec-morph}. This comparison of photometry and fitted SEDs also serves as further validation that the GALEX {\em NUV} magnitudes for A2063 are correct, and that the {\em UVM2} photometry for this cluster needs to be adjusted to reflect the proper integration time.

In the cases of {\em UVM2} and {\em UVW1}, the predicted magnitudes will differ from the measured ones by any unknown offset in our calibration (i.e., assumed zero point magnitude) and by the extinction. Ideally, we would thereby take the vectors of measured minus predicted magnitudes along with the $E(B-V)$ values and perform a simple linear least squares fit to determine this offset and the appropriate $R_X$ to translate $E(B-V)$ into extinction. However, as previously indicated our clusters have a very small range in $E(B-V)$ especially when one considers that only one galaxy from A400, the sole cluster with $E(B-V) > 0.05$, has complete photometry in the seven GALEX+SDSS bands. Thus, the spread in $E(B-V)$ is too small to produce a meaningful estimate of $R_X$ and we instead investigate individual pairings of zero point offset and $R_X$ in comparison with the GALEX and SDSS filters. We find that when using $R_{UVM2} = 8.2$ (i.e., the same value as that for {\em NUV}), the dispersion in the difference between measured and predicted magnitudes is 0.365 and near its minimum. As expected, changing the value of $R_{UVM2}$ has little effect ($\le 0.002$ in the measured dispersion) on the results for $6.8 \leq R_{UVM2} \leq 9.2$. We have also used the {\em UVM2} filter curve itself along with the IDL routine {\ttfamily dust\_intfilter} to directly determine $R_{UVM2}$ for the ``average'' SDSS galaxy spectrum in the direction of A1656 and obtained 8.8. The negative mean in the difference between measured and predicted magnitudes (i.e., measured magnitudes are slightly brighter than those predicted) is opposite in direction to the one seen when directly comparing {\em UVM2} to GALEX {\em NUV} magnitudes (Figure \ref{fig-galexck}). Our preliminary approach is thus to keep the zero point magnitude unchanged and use $R_{UVM2} = 8.2$. Similarly, for {\em UVW1} we find that the adopted zero point magnitude is consistent with the results for the other filters and $5.5 \leq R_{UVW1} \leq 6.3$. In this case, {\ttfamily dust\_intfilter} returns $R_{UVW1} = 6.0$ and we subsequently adopt this value.

\section{Application}\label{sec-apply}

\subsection{Color Magnitude Diagrams for Galaxy Clusters}\label{sec-cmd}

\begin{figure}
\figurenum{7}
\epsscale{0.9}
\plotone{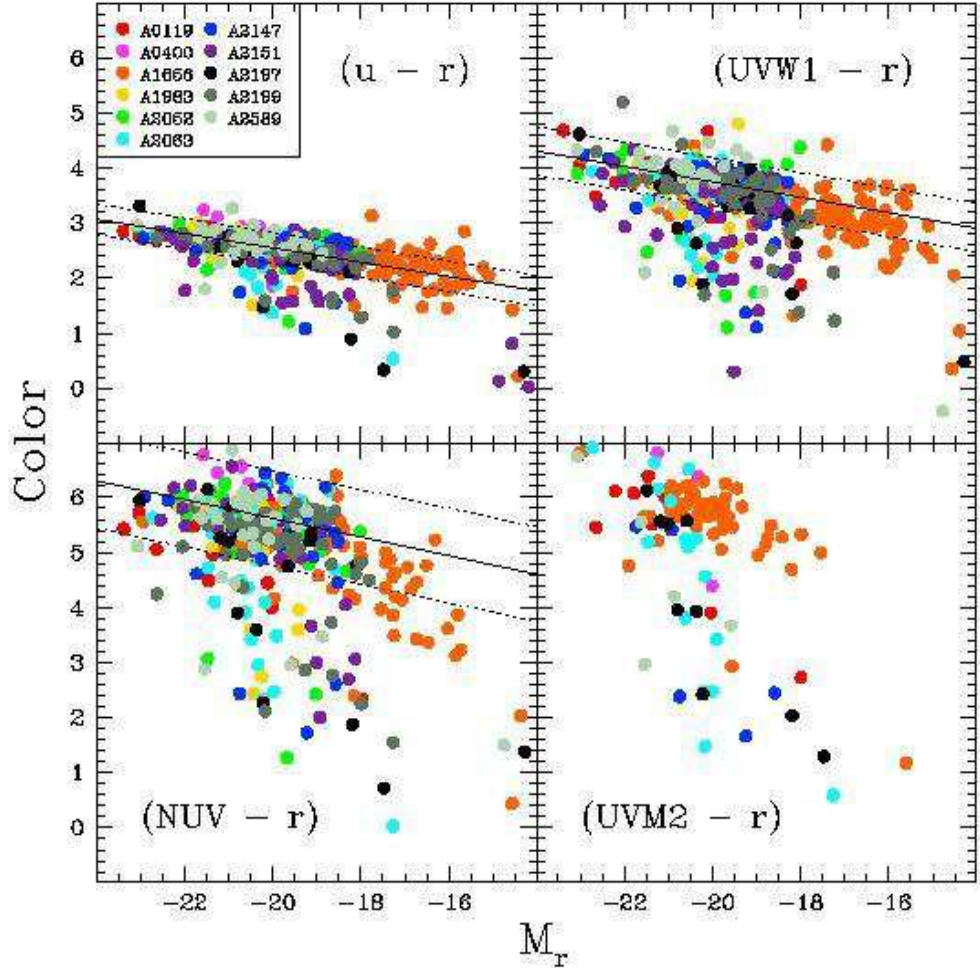}
\caption{Color magnitude diagrams (CMDs) for galaxies with spectroscopic redshifts confirming cluster membership. Each panel is labeled showing the pair of filters that have been used to evaluate the colors, and the top left panel includes the key for color-coding of the points by galaxy cluster. Red sequence fits (see text) are shown by solid lines, with dashed lines indicating the 1$\sigma$ range about the fitted relations. No fit is provided for ($UVM2 - r$) as the smaller sample size prevented the fitting routine from converging, although the similarity between the $UVM2$ and $NUV$ filters provides perspective.\label{fig-cmds}}
\end{figure}

Motivated by the numerous GALEX studies, we have explored color magnitude diagrams (CMDs) constructed using photometry for galaxies with measured redshifts that place them within the eleven clusters. Figure \ref{fig-cmds} shows four such CMDs, in each case with color plotted against absolute magnitude in SDSS {\em r}. K corrections determined from the \kcorrect routine have been applied to shift all photometry points to $z=0$, although these corrections are small, $\lesssim 0.1$ mag in {\em r} and $\lesssim 0.2$ mag in the UV filters, on account of the low redshifts of the sample. For the CMD based only on SDSS data, these k corrections were determined using only the five filter SDSS photometry, while for the XMMOM/optical colors we used the SDSS plus the {\em UVM2} and {\em UVW1} filters and for the GALEX/SDSS CMD we used the SDSS plus {\em FUV} and {\em NUV} filters. The {\em UVM2} magnitudes for A2063 have been adjusted by 0.81 mag to reflect the correction for integration time, and the magnitude errors arbitrarily increased by 0.1 magnitudes to reflect their additional uncertainty when using them to estimate K corrections. In each diagram, we fit the color magnitude relation using the procedure of \citet{lopezcruz2004} although we differ from that work in that we do not use a cut in radial distance from the cluster centers \citep[the XMM-OM images are smaller than the areas contained by $r_{200}$, the cut used in ][]{lopezcruz2004}, and in that we determine the fit using galaxies with measured redshifts as opposed to using the photometry for all objects with non-stellar profiles. Using only galaxies with cluster redshifts provides more accurate fits to the color magnitude relations, and these will be applied later to the categorization of galaxies having only photometric data. The brief summary of the fitting procedure is that the color-magnitude relation is fit as a straight line for all galaxies with $-18 \le M_r \le -22$ (chosen as the range where the red sequence can be seen to be the dominant population for all colors). We find the slope and intercept that minimize the difference between the actual and fitted colors of the data through the biweight location and scale of this distribution \citep{beers1990}. The use of these robust statistics renders the fit less affected by outliers, which in this case are those cluster galaxies with active star formation or AGN producing blue colors. The fitting is iterative, with data more than 3$\sigma$ removed from the fitted relation dropped and a subsequent fit determined. For all but {\em UVM2}, this quickly converges to a stable solution where subsequent iterations do not reject any objects. The paucity of {\em UVM2} data points makes fitting the relation using such colors ineffective, but the rough equivalence of the {\em UVM2} filter to the GALEX {\em NUV} filter provides perspective. Our fitted relations are: $(u - r) = -0.05 - 0.13M_r$ (for $N=365$ galaxies in the fit), $(UVW1 - r) = 0.95 - 0.14M_r$ ($N=273$), and $(NUV - r) = 2.22 - 0.17M_r$ ($N=273$). We note that the relation for $(NUV - r)$ is consistent with that reported in \citet{wyder2007} after passive evolution between the $z=0.1$ of that study and the $z=0$ used here.



\begin{figure}
\figurenum{8}
\epsscale{0.9}
\plotone{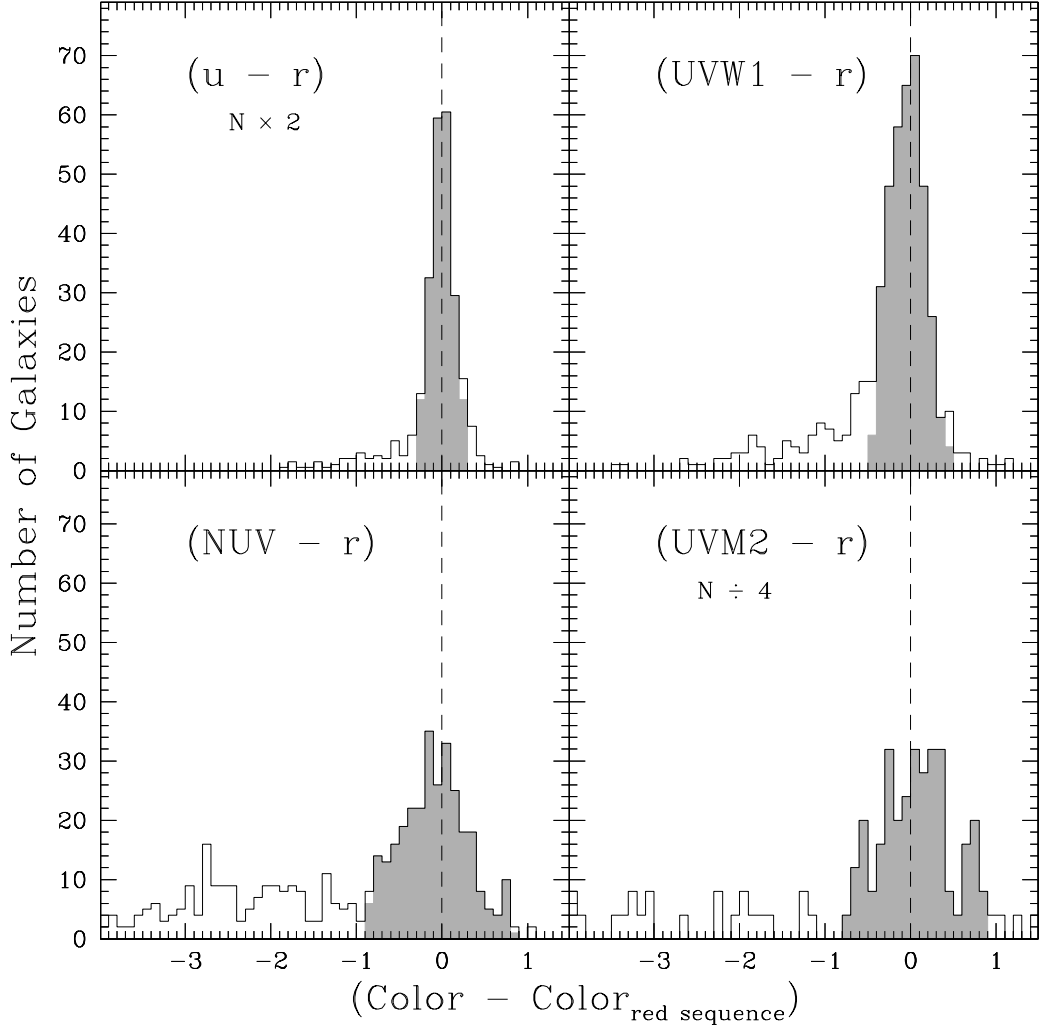}
\caption{Number of galaxies by color relative to the fitted color magnitude relation (see Figure \ref{fig-cmds}). The relative color is defined as the actual measured color minus that predicted by the fitted color magnitude relation for a galaxy with the same absolute magnitude. Negative values thereby indicate galaxies bluer than the red sequence. Galaxies within $\pm 2\sigma$ of the fitted color magnitude relation are indicated as the shaded portion of the histogram. For {\em UVM2}, where low numbers make the red sequence fit problematic, we have adopted the ($UVM2 - r$) fitted red sequence. The number of galaxies in the ($u - r$) panel has been multiplied by 2 (i.e., a value of 50 on the y axis indicates 100 galaxies) while that in the ($UVM2 - r$) panel has been divided by 4 in order to assist in comparing the panels. In all panels, a dashed line marks a color offset of zero: the location of the fitted color magnitude relation.\label{fig-colhists}}
\end{figure}

It can be seen from Figure \ref{fig-cmds} that the cluster galaxies of our study are dominated by objects on the red sequence, as would be expected based on the morphology-density relation and the XMM-OM data sampling the cluster cores. This general lack of star forming galaxies prevents us from decomposing the color information into separate red and blue sequences and we subsequently divide red and blue galaxies using a 2$\sigma$ deviation from the fitted red sequence. There is a much greater spread in the GALEX/SDSS colors than for colors determined solely from SDSS data, both in total and for galaxies lying along the red sequence. This has been discussed in many prior studies and reflects the ability of near-UV data to probe recent star formation and thus better assess star formation history \citep[e.g.,][]{wyder2007,kaviraj2007}. We might also gain this information from the {\em UVM2} filter as it is roughly the same as {\em NUV}, although within the present study we lack sufficient numbers of {\em UVM2} detections. However, we do have ample numbers of galaxies with {\em UVW1} photometry and the same general effect is apparent. To further illustrate this, in Figure \ref{fig-colhists} we plot histograms of the colors of galaxies relative to their fitted red sequences for the same color pairings as were used in Figure \ref{fig-cmds}. These histograms show clearly that the width of the red sequence increases from $(u - r)$ to $(UVW1 - r)$ to $(NUV - r)$, with dispersions of 0.14, 0.22, and 0.43, respectively. These values reflect both real variation in galaxy colors about the red sequence and measurement errors of the data, with the median photometric errors being 0.06, 0.06, and 0.18 for $u$, $UVW1$, and $NUV$. It is also apparent that the tail of the distribution -- galaxies significantly bluer than the red sequence -- becomes more populous as UV information is included. To this end, there are 57 cluster galaxies for which $(UVW1 - r)$ colors suggest a greater than 2$\sigma$ separation from the red sequence whereas their $(u - r)$ colors are consistent with the red sequence at this level. 

\begin{figure}
\figurenum{9}
\epsscale{0.9}
\plotone{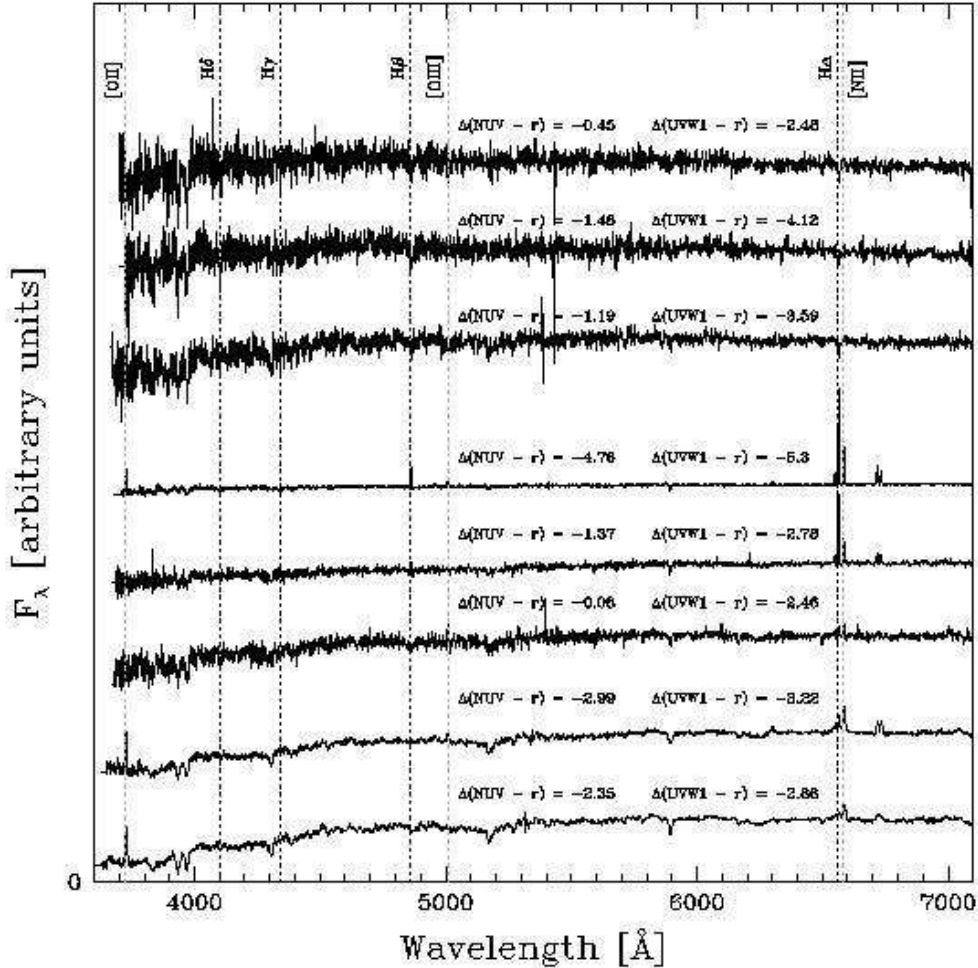}
\caption{SDSS spectra for objects with unflagged {\em NUV} and {\em UVW1} photometry and ($UVW1 - r$) colors more than 2$\sigma$ off the red sequence yet consistent with the ($u - r$) red sequence to within 2$\sigma$. All spectra have been shifted to the rest frame and the location of diagnostic features (Balmer lines, [{\scshape O~ii}], [{\scshape O~iii}], and [{\scshape N~ii}]) are indicated by vertical dotted lines. The statistical significance of each galaxy's deviation from the red sequence is indicated directly above each spectrum. The bottom three galaxies are low-luminosity AGN, the middle two are star-forming galaxies, and the top three appear to be weak post-starbursts. The post-starbursts are arranged in order of the strength of their H$\delta$ absorption, with the uppermost galaxy having an equivalent width of $5.2 \pm 1.2$ $\mbox{\AA}$ and the lower two having equivalent widths of $2.7 \pm 0.6$ $\mbox{\AA}$ and $1.7 \pm 0.4$ $\mbox{\AA}$, respectively.\label{fig-sdspec}}
\end{figure}

What can be said about these 57 galaxies with blue $(UVW1 - r)$ colors but $(u - r)$ colors consistent with the red sequence? Twenty-six of them do have flags on their {\em UVW1} photometry indicating that the source was near the edge of an image or was deblended from nearby neighbors. The remaining 31 galaxies include 18 with GALEX {\em NUV} photometry and 20 with SDSS spectra, with eight galaxies having both. For half of the objects with GALEX photometry the $(NUV - r)$ colors are consistent (at 2$\sigma$) with the red sequence defined using this color, while the other half agree with the $(UVW1 - r)$ impression of blue colors. In each case where the $(NUV - r)$ colors do not show a significant deviation from the red sequence the {\em NUV} data come from the AIS and thus have large associated errors at least partially explaining this lack of significant blue colors. We plot the eight available SDSS spectra for objects with GALEX photometry and discrepant SDSS and XMMOM-SDSS colors in Figure \ref{fig-sdspec}, along with their separations from the $(NUV - r)$-defined and $(UVW1 - r)$-defined red sequences. At least two appear to be AGN consisting of an underlying old stellar population plus weak emission lines with [{\scshape N~ii}]$\lambda 6584$ stronger than H$\alpha$, plus relatively strong [{\scshape S~ii}]$\lambda\lambda 6717+6731$ and [{\scshape O~ii}]$\lambda3727$. For each of these, the $(NUV - r)$ and $(UVW1 - r)$ colors agree that the galaxies are slightly off the red sequence. A third galaxy is likely in the same class but is a fainter object with a correspondingly noisier spectrum. Two galaxies appear to be star-forming galaxies on the basis of their emission lines and standard line ratio diagnostics. The remaining three galaxies are faint and have noisy spectra, but in each case Balmer absorption features are notable. The SDSS indicates equivalent widths of $5.2 \pm 1.2$, $2.7 \pm 0.6$, and $1.7 \pm 0.4$ $\mbox{\AA}$ for the H$\delta$ absorption line for these three galaxies, consistent with the recent cessation of star formation and thus a general post-starburst classification.

Conversely, only four galaxies have $(UVW1 - r)$ colors consistent with the red sequence while their $(u - r)$ colors suggest a greater than $2\sigma$ separation from the red sequence. One of these four galaxies has a flag on its {\em UVW1} photometry, and in each case the $(NUV - r)$ colors agree with the depiction gleaned from the {\em UVW1} data. When considered along with the $(UVW1 - r)$-defined blue galaxies, these findings argue that $(UVW1 - r)$ colors provide an accurate depiction of galaxy activity, and frequently this depiction improves upon that garnered from SDSS data alone.

\begin{figure}
\figurenum{10}
\epsscale{0.9}
\plotone{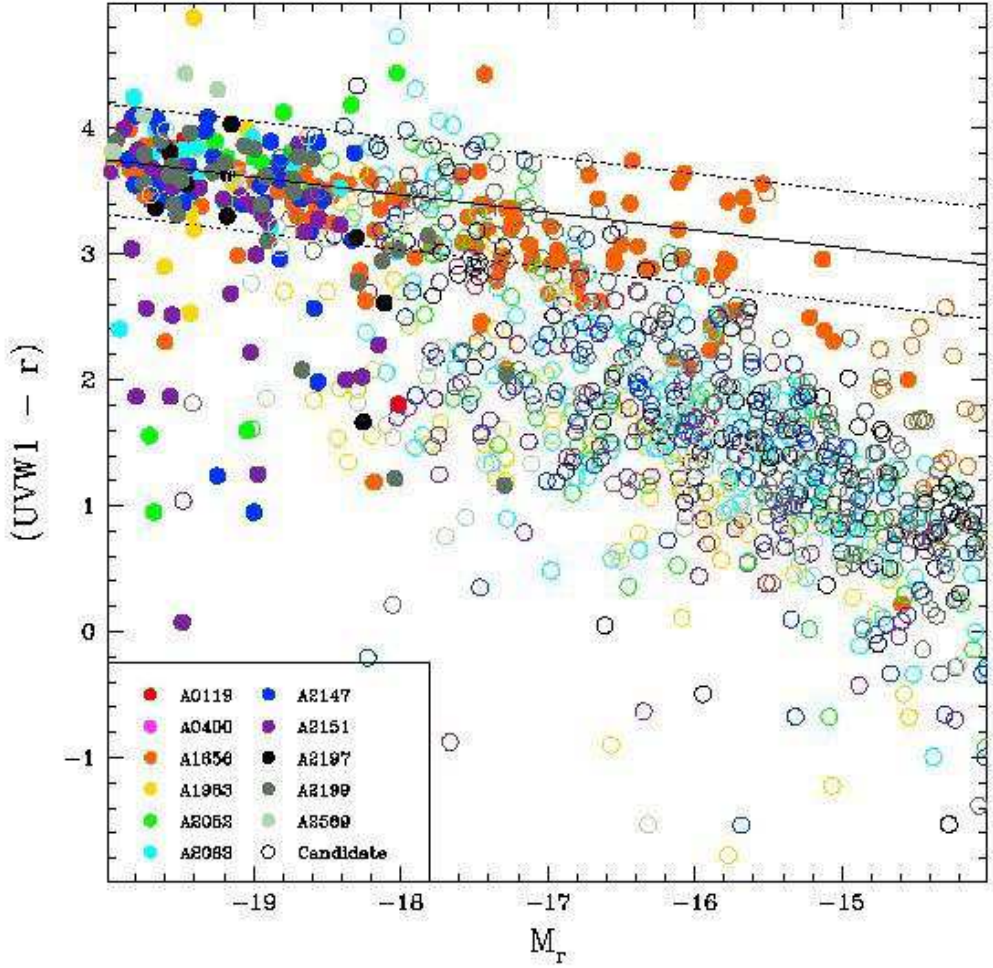}
\caption{Color magnitude diagram constructed using {\em UVW1} data, with spectroscopically confirmed cluster members indicated by solid points and galaxies without measured redshifts indicated by open points. The red sequence fit is shown by a solid line, with its $\pm2\sigma$ range given by dotted lines. The lack of points toward the top right is caused by the shallower {\em UVW1} data precluding the detection of faint red objects.\label{fig-cmdphot}}
\end{figure}

As a corollary to this, the XMM-OM data are valuable for photometric selection of galaxies. For cluster studies, they accurately identify the red sequence and galaxies significantly more red than this may be assumed to be background galaxies to limit the number of targets for spectroscopic follow-up. Figure \ref{fig-cmdphot} shows the fainter end of the $(UVW1 - r)$ vs. {\em r} CMD for both spectroscopically-confirmed cluster galaxies and all galaxies without spectroscopic redshifts. In general, our clusters are well sampled spectroscopically and the brightest possible cluster galaxy without a measured redshift would have $M_r \approx -20$ should it reside in its associated cluster. This is largely a result of the SDSS Main Galaxy Sample providing spectra for nearly all galaxies down to {\em r} of about 17.8, or roughly $M_r = -18.7$ for our most distant cluster. Below this absolute magnitude there are many candidate cluster member galaxies, although strong selection biases are at work. The apparent decline in candidate red sequence galaxies is caused by the shallow depth of the {\em UVW1} images relative to the SDSS {\em r} data, as redder colors fall below the detection threshhold of the {\em UVW1} data. For A1656 where a deeper {\em UVW1} image is available thanks to the combination of multiple observations, red sequence galaxies continue to be detected down to $M_r \gtrsim -16$. There appear to be many candidate faint star-forming galaxies with $(UVW1 - r)$ between about 0.5 and 2.5 and $M_r \gtrsim -17$, although the majority of these are likely to be background galaxies. This assessment is based on A1656, where spectroscopic surveys have probed to very faint populations and we find that nearly all of the galaxies in this range of color and absolute magnitude are background galaxies and hence not cluster members.

Ideally, we would like to use {\em UVW1} data in conjunction with optical data to identify ``green valley'' objects. We are precluded from definitively doing this by the low numbers of blue galaxies and the corresponding inability to fit the blue sequence. However, we anticipate that future studies of non-cluster selected fields with {\em UVW1} data will be able to accomplish this objective in a manner directly analogous to that for $(NUV - r)$ work and will explore this in a future paper.

\subsection{Morphological Dependence}\label{sec-morph}

\citet{schiminovich2007} investigated how different galaxy morphologies related to regions of the UV-optical CMD, motivated by the interpretation of CMDs as parameterizations of star formation history and total stellar mass. They found a ``star-forming sequence'' along which galaxies have nearly constant star formation rate surface density, with disk-dominated galaxies tightly following this sequence and bulge-dominated galaxies exhibiting a much larger spread in star formation rate for a given mass. Here, we use our much smaller sample to do a preliminary look at how our findings are related to galaxy morphology.

The Galaxy Zoo project provides visual classifications of many galaxies from the SDSS \citep{lintott2008}. Hundreds of thousands of volunteers have inspected SDSS galaxy images to place them within the general morphological categories of elliptical, spiral (with three subcategories pertaining the clockwise/anti-clockwise winding of their arms or ``other'' including edge-on orientation), mergers, and ``Don't Know.'' After careful statistical evaluation of the classifications compared within the collection of volunteers and against professional and automated classifications, the Galaxy Zoo morphologies were presented in a published catalog \citep{lintott2011}. We have used the Galaxy Zoo classifications that remove known sources of bias and provide ``clean'' morphologies, where clean refers to an agreement on classification of at least 80\% of participants after correction for known biases.

As expected, effectively all galaxies with elliptical morphologies reside on the red sequence in each of the color pairings we examined in the previous discussion. Four elliptical galaxies have apparent blue colors in either $(UVW1 - r)$ or $(NUV - r)$ (two for each). In each case only one of the two colors places the object off the red sequence, and these data have flags on the relevant UV photometry whereas the other UV photometry is unflagged and places the object on the red sequence. Similarly, most of the spirals lie off the red sequence but there are a few exceptions. Seven spiral galaxies have colors placing them on the red sequence, with the three color pairings generally producing consistent interpretations. Three of these have SDSS spectra with emission lines indicating an active nucleus, and another two have absorption-line spectra with no evidence for emission and are thus ``passive spirals'' \citep[e.g.,][]{dressler1999}. The remaining two were consistent with the red sequence for $(u - r)$ but not for $(UVW1 - r)$, and were thus included in Figure \ref{fig-sdspec}. They are the uppermost star-forming galaxy (i.e., 4th galaxy from the top) and the middle post-starburst galaxy (second galaxy from the top) in that Figure. The former is only located on the red sequence for $(u - r)$ and may indicate an error in its $u$-band photometry.

We also revisited the photometry and \kcorrect{} fits described in Section \ref{fig-extinct}, and include the results in Table \ref{tbl-ecal}. Despite the relatively small numbers of sources with full GALEX and SDSS photometry plus Galaxy Zoo classifications, we find a significant difference between elliptical and spiral galaxies in the \kcorrect{} fits to their UV spectra. In general, the fitted SEDs accurately match both the optical and UV photometry for the spiral galaxies. This is also true for the galaxies with ``uncertain'' classifications, which are sources for which neither the elliptical nor spiral classification received the requisite 80\% agreement threshold for a clean classification. However, the UV magnitudes predicted by the SED fits for elliptical galaxies are significantly fainter than the actual measured magnitudes for such sources. At both $FUV$ and $NUV$, this offset is significant at roughly 5$\sigma$. This may result from the presence of the ``UV upturn'' in elliptical galaxies \citep[see review by][]{oconnell1999}, which is the rise in the flux density of many elliptical galaxies and spiral bulges at wavelengths $\lesssim2500\mbox{\AA}$. At the low redshifts of our sample, the UV upturn falls primarily within the $FUV$ band but extends slightly into $NUV$ and $UVM2$. The UV upturn does not coincide with the $UVW1$ filter and indeed we do not see any significant difference between predicted and measured magnitudes for elliptical galaxies within this filter. Potential issues caused by the lack of extreme horizontal branch stars and the UV upturn in the synthesis models used by \kcorrect{} are specifically noted in that work, and thus the presence of the UV upturn in actual elliptical galaxies would make their measured magnitudes brighter than those predicted by the code as we observe. The strength of the UV upturn varies significantly among galaxies although is generally stronger in more metal-rich and brighter elliptical galaxies. We do find this to be consistent with our findings in that the difference between measured and predicted magnitudes is usually larger for brighter galaxies although we do not find this to be statistically significant.

\subsection{Star Formation Histories of Cluster Galaxies}\label{sec-sfh}

\begin{figure}
\figurenum{11}
\epsscale{0.9}
\plotone{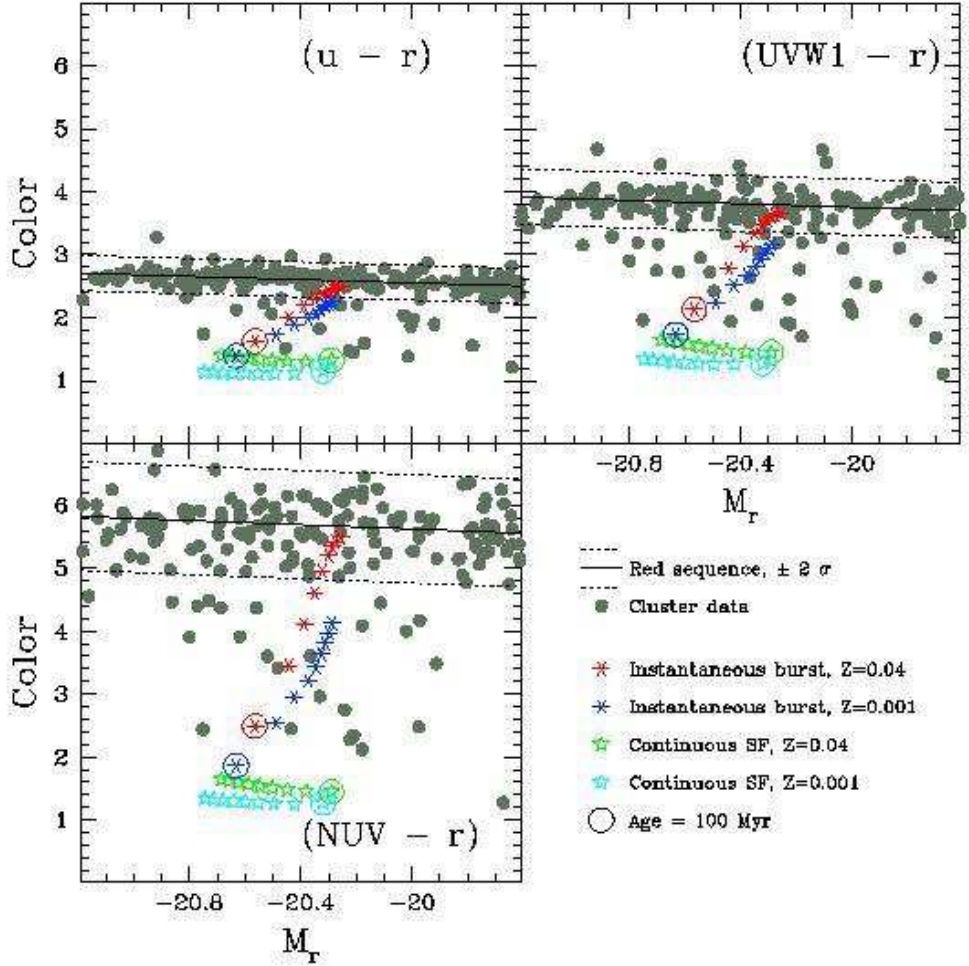}
\caption{Behavior of toy model star-forming galaxies on color magnitude diagrams (CMDs). Actual data from our sample are shown as grey filled circles, with fitted red sequences and their 2$\sigma$ limits shown with solid and dashed lines, respectively (refer to Figure \ref{fig-cmds}). Model galaxies are depicted with asterisks (instantaneous starburst) and stars (continuous star formation), where their evolution is followed from 100 Myr (indicated by the large circled asterisk) to 900 Myr.\label{fig-sfhmodel}}
\end{figure}

We can gain additional perspective on the CMDs and their relation to star formation history by examining simple toy models. We start with a galaxy residing on the red sequence at $M_r = -20$ and then add star formation using the Starburst99 synthesis models \citep{leitherer1999}. In one model, we assume the galaxy undergoes an instantaneous burst of star formation that increases its stellar mass by 25\% and then follow its color evolution over the subsequent 900 million years. In a second model, we assume the galaxy undergoes constant star formation at a rate of 1 M$_\odot$ yr$^{-1}$. In this latter model, after 900 million years the galaxy has increased its stellar mass by just over 30\% and is thus comparable to the instantaneous burst model. For both the instantaneous burst and constant star formation cases we use Starburst99 models for each low and high metallicity ($Z = 0.001$ and $Z = 0.04$) to bracket the possible range. The resulting tracks are shown in Figure \ref{fig-sfhmodel} plotted along with actual data for comparison.

The general behavior of the models is the same for each color pair. The instantaneous starburst models lie at the edge of the outer envelope of observed blue galaxies shortly after their burst, then fade and redden as the massive stars formed in their starbursts die off. If the metallicity of the stars formed in the starburst is low enough, the galaxy remains bluer than the red sequence even after 900 million years for each of the color pairings. For higher metallicity bursts, the galaxy colors become consistent with the observed red sequence after about 400 million years for each $(u - r)$ and $(UVW1 - r)$ and slightly longer for $(NUV - r)$. Continuous star formation models maintain roughly constant colors while gradually becoming brighter as they build up stellar mass.

\begin{figure}
\figurenum{12}
\epsscale{0.9}
\plotone{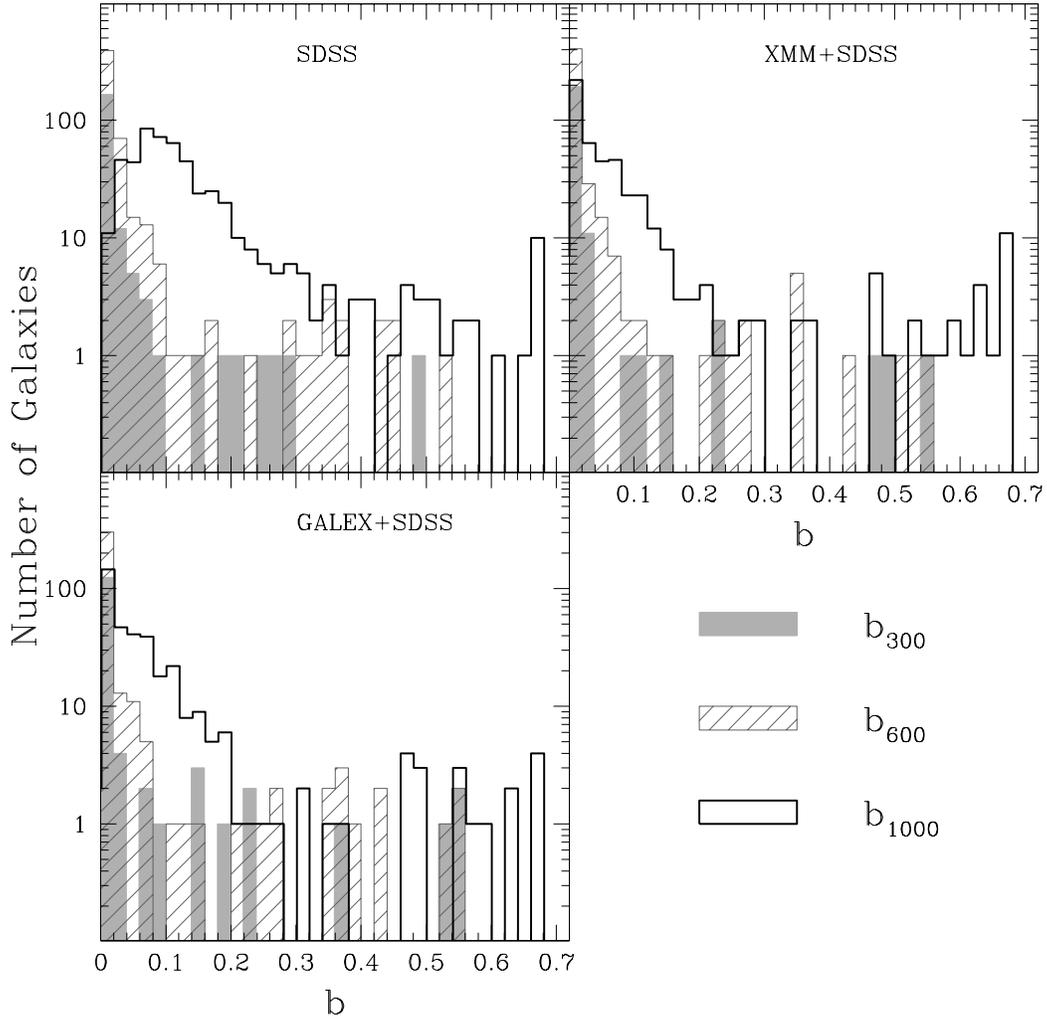}
\caption{Histograms of number of cluster galaxies at given values of $b_X$, where $X$ is 300 Myr (grey filled histograms), 600 Myr (hatched histograms), and 1000 Myr (solid black histograms). The $b$ values are determined using \kcorrect with input being SDSS photometry only (top left), SDSS plus XMM-OM (top right), and SDSS plus GALEX (bottom left). Note how the addition of ultraviolet data greatly reduces the spread in $b_{1000}$.\label{fig-sfhhist}}
\end{figure}

As noted earlier, \kcorrect{} provides a means to evaluate star formation histories by decomposing the populations that produced the best-fitting template. A simple cut at this information is provided by the stellar birthrate parameter, $b_X$, the ratio of the stellar mass formed within the past $X$ years to that formed over the history of the universe for a given galaxy \citep[a variant on the ratio of the current SFR to the past-averaged SFR initially used by][]{kennicutt1983}. We have used \kcorrect{} to estimate $b_{300}$, $b_{600}$, and $b_{1000}$, where the subscripts indicate the stellar mass formed within that past number of Myr. While we might expect the addition of ultraviolet data to reveal recent star formation missed by optical-only photometry, it actually provides stronger contraints on the lack of star formation in cluster galaxies within the past billion years. This can be seen in Figure \ref{fig-sfhhist}, which shows histograms of the $b_X$ values for the cluster galaxies of our sample. Using only SDSS data to estimate star formation histories (top left panel), there is a broad distribution in $b_{1000}$ with nearly 98\% of all galaxies (507/518) having $b_{1000} > 0.02$ and about 50\% (260/518) having $b_{1000} > 0.10$. The addition of XMM-OM data to the SDSS photometry reduces these figures to 54\% (271/499; we use only those galaxies with {\em UVW1} data and hence the sample is slightly smaller than that based on SDSS photometry alone) and 19\% (93/499), respectively, and the same figures based on GALEX and SDSS data are 60\% (223/371, as 371 galaxies have both SDSS and GALEX {\em NUV} photometry) and 21\% (78/371). The ultraviolet data, be they from XMM-OM or GALEX, thereby rule out significant amounts of star formation for most of our cluster galaxies over the past billion years. The difference is less on shorter timescales where the signatures of current star formation are stronger and more easily obtained even without ultraviolet data. The SDSS data alone find only about 6\% (29/518) of our galaxies show any evidence for star formation over the past 300 million years, with $b_{300} > 0.02$. Finding large amounts of recent star formation is even more rare in our cluster sample, with only 8 galaxies (less than 2\% of the total) having $b_{300} > 0.10$. These results are consistent with those determined when XMM-OM and GALEX data are included. The XMM-OM data in conjunction with the SDSS data find about 4\% (20/499) of galaxies have $b_{300} > 0.02$ and less than 2\% (8/499) have $b_{300} > 0.10$. For GALEX, these numbers are about 5\% (19/371) and 3\% (12/371). The overlap between the sources having $b_{300}$ based on the different combinations of photometric bands is strong, and any discrepancies are understood as slight shifts in the timescale of recent star formation (for example, if the SDSS data showed $b_{300} > 0.02$ whereas the GALEX and SDSS data combined showed $b_{300} < 0.02$, the GALEX and SDSS data would find $b_{600}$ several times larger than 0.02).

In Section \ref{sec-cmd} and Figure \ref{fig-sdspec} we discussed several objects with discrepant colors. Three of these were suggested to have post-starburst characteristics, and we can assess their $b_X$ values in light of this. The two galaxies with the stronger H$\delta$-absorption have $b_{300} = 0.00$ yet have $b_{600} \approx 0.05$ and $b_{1000} \approx 0.06$, consistent with the weak post-starburst classification. The addition of UV data does not greatly alter this interpretation and serves only to reduce slightly the magnitude of the recent star formation: $b_{1000}$ based on SDSS data only is about 0.07 while that including near-UV data is about 0.06. The results are similar for the third of these galaxies that had the weakest H$\delta$ absorption, although its recent star formation is a lower fraction of its total stellar mass.

\subsection{Photometric Redshifts and Stellar Masses}\label{sec-pzandm}

Both large area surveys with too many objects for spectroscopic follow-up and the importance of selected area deep fields with objects too faint for practical spectroscopic observation have made phometric estimation of redshifts a heavily-used and important technique. We have investigated the application of XMM-OM photometry for photometric redshift determination by using the photo-z tools provided by \kcorrect{} in order to determine whether the inclusion of XMM-OM photometry provides noticeable improvement in the accuracy of photometric redshifts. We evaluate the accuracy of the photometric redshifts by looking at the difference between the photometric and spectroscopic redshifts $(z_{phot} - z_{spec})/(1 + z_{spec})$, where we implicitly assume the spectroscopic redshifts are accurate. The dispersion in this quantity, $\sigma_z$, represents the accuracy of the photometric redshifts. A single round of 3$\sigma$ clipping was used to remove a small number of ``catastrophic failures'' where the photometric redshifts are severe outliers to the overall trend, and we evaluated only those objects for which the near-UV photometry ($FUV$, $NUV$, $UVM2$, and $UVW1$) had errors smaller than 0.5 magnitudes. The results may be found in Table \ref{tbl-photoz}.

In general, the addition of UV photometry does little to improve photometric redshifts as those derived using the SDSS photometry alone are of comparable quality to those determined when UV filters have been added to the SDSS photometry. This might be expected, as the strongest feature for photometric redshift estimation in the near-UV/optical portion of the spectrum is the 4000-Angstrom break, which is already sampled by the $u$ and $g$ photometric bands. As one moves to higher redshift, the addition of the UV filters is of even lesser importance as the 4000-Angstrom break shifts into redder filters where the associated SDSS photometric errors are smaller than those for the SDSS $u$ photometry. We confirm this by separating our sample into cluster galaxies (as before, those within $z \pm 0.01$ of the cluster systemic velocity) and background galaxies (those with $z > z_{sys} + 0.01$). The typical error in photometric redshifts for the low redshifts of our cluster sample is $\sigma_z \approx 0.05$, while that for higher redshifts is $\sigma_z \approx 0.1$. At the lower redshifts of the cluster sample there is marginal improvement in the accuracy of the photometric redshifts that include the near-UV data. This is most important for fainter galaxies where the photometric errors in $u$ are relatively large. For sources with $u > 20$ and $z_{spec} < 0.1$, we find that substitution of $UVW1$ for $u$ photometry (and hence estimation of photometric redshifts using $UVW1~g~r~i~z$ filters) improves $\sigma_z$ from 0.11 to 0.05.

\begin{figure}
\figurenum{13}
\epsscale{0.9}
\plotone{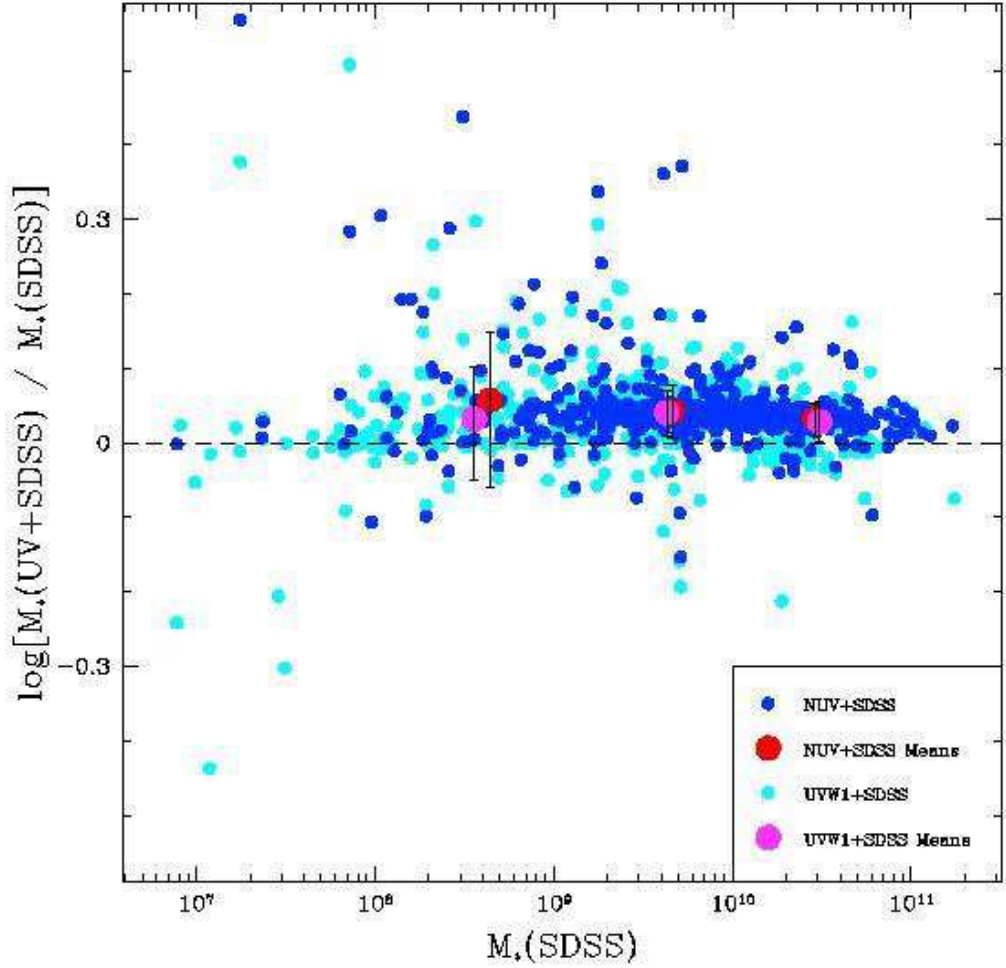}
\caption{Comparison of stellar masses estimated using near-UV data (GALEX $NUV$ and XMM-OM $UVW1$) with those estimated using only SDSS photometry. The points indicating means have been evaluated for sources with SDSS masses below $10^9$M$_\odot$, between $10^9$M$_\odot$ and $10^{10}$M$_\odot$, and above $10^{10}$M$_\odot$.\label{fig-masses}}
\end{figure}

Prior studies have shown that the addition of UV data can improve the accuracy of stellar mass estimates \citep{salim2005}. Without knowing the true stellar masses of galaxies in our sample, we can again use stellar masses estimated using only SDSS photometry as a reference for those estimated after adding UV data to the SDSS photometry. We find that \kcorrect{} suggests a mass increase of about 10\% when UV data are included, ranging from about 8\% when only $UVW1$ are added to almost 12\% when GALEX $NUV$ and $FUV$ are included, consistent with the findings reported by \citet{salim2005}. This indicates that even the addition of one photometric measurement blueward of SDSS $u$ is valuable in estimation of stellar mass (see Figure \ref{fig-masses}). Within our sample, the increase appears dependent on galaxy mass with more massive galaxies having smaller relative increases. For galaxies with SDSS-derived stellar masses above $10^{10}$M$_\odot$, the addition of GALEX photometry produces an average increase in the estimated stellar mass of about 8\%. Conversely, stellar masses derived using GALEX plus SDSS photometry are nearly 33\% larger than their values derived using only SDSS photometry when the SDSS-only stellar masses are $<10^9$M$_\odot$. This result is at least in part a result of our object selection, as we require a detection in XMM-OM UV filters and fainter red sequence objects will often have $UVM2$ and $UVW1$ magnitudes falling below our detection threshold.


\subsection{Color-Color Diagrams and Non-Cluster Object Characterization}\label{sec-nonclus}

\begin{figure}
\figurenum{14}
\epsscale{0.9}
\plotone{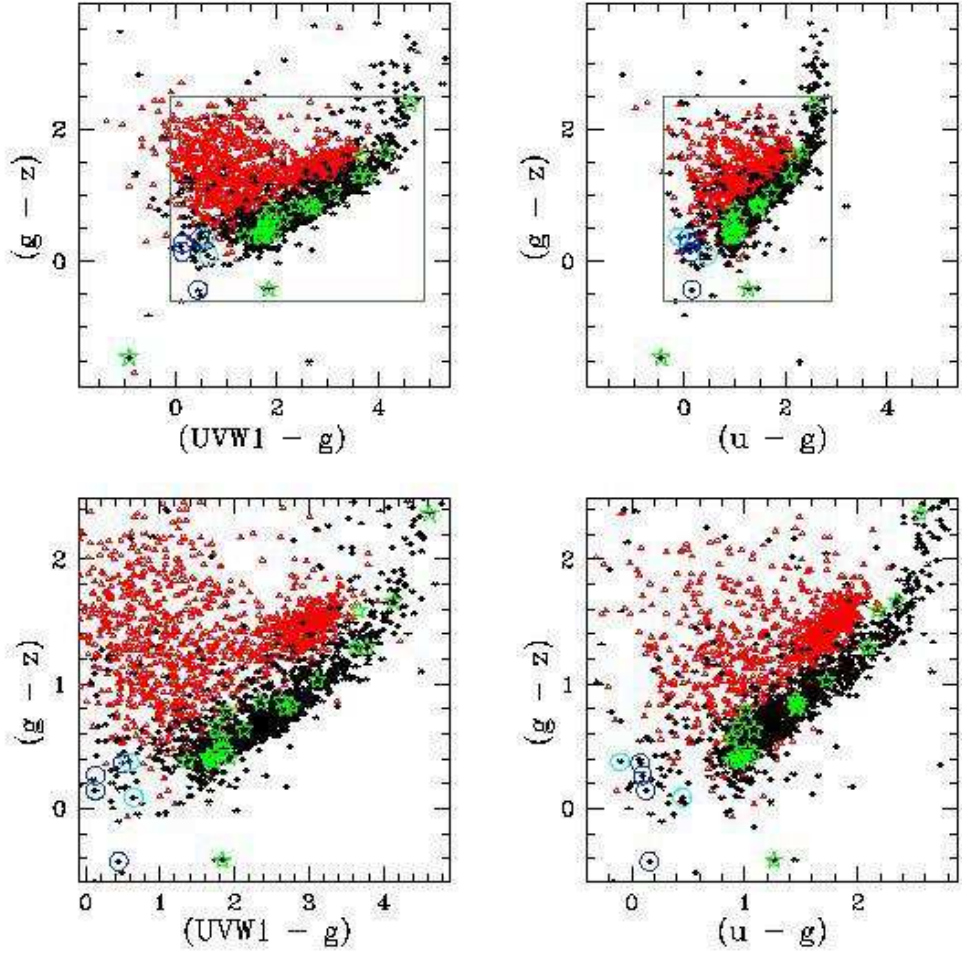}
\caption{Color-color plots based on {\em UVW1} and SDSS photometry (left) and SDSS only photometry (right). Objects with large errors in their magnitudes ($> 0.3$) are excluded, as are objects with $m_r < 13$ to help remove stars with incorrect photometry due to saturation. The different source types are indicated by symbols: stars classified by the SDSS photometric pipeline (black asterisks), stars with SDSS spectroscopy (large green stars), SDSS photometric galaxies (grey triangles), spectroscopically-confirmed cluster galaxies (red triangles), and quasars (blue open circles, with dark blue being SDSS spectroscopic quasars and cyan being quasars with NED redshifts). The top panels are plotted with the same limits on the axes to capture the majority of the data and show the difference in spread between {\em UVW1} and $u$, while the bottom panels are zoomed in on the stellar sequence (regions depicted by grey boxes in top panels) to better show the cleaner separation of source types with the addition of {\em UVW1} data.\label{fig-colors}}
\end{figure}

Colors utilizing {\em UVW1} data are also advantageous to selecting different classes of objects. \citet{seibert2005} showed this using GALEX {\em NUV} along with SDSS optical filters in color-color plots. They were able to segregate many different populations of stars (main sequence, white dwarfs, M dwarfs, etc.) and found improved separation of galaxies and stars particularly because the GALEX MIS {\em NUV} data are deeper than SDSS {\em u} data. We demonstrate this for XMM-OM {\em UVW1} in Figure \ref{fig-colors}, which compares $(UVW1 - g)$ and $(u - g)$ color plotted against $(g - z)$ color. These filters have been chosen to give large spreads in colors and hence more separation of object types in color-color space. Stars and galaxies defined on the basis of the SDSS photometry are indicated by black asterisks and grey open triangles, respectively. A well-defined sequence of normal stars is seen in both plots \citep[see, for example, ][]{newberg1999,covey2007}, although the larger range in $(UVW1 - r)$ color appears to provide a cleaner separation between stars and galaxies. The nearby cluster galaxies investigated in this work are shown with red open triangles, and are strongly bunched in color-color space as they are predominantly red sequence objects and by selection are all at low redshift. The six objects with $z > 1$ are indicated by blue open circles, with those having SDSS spectra being dark blue and those with NED redshifts being cyan. One of the SDSS high redshift objects (at 12$^{\mbox{{\scriptsize h}}}$59$^{\mbox{{\scriptsize m}}}$34\fs6 $+$27$^\circ$57\arcmin53\arcsec ) has an uncertain classification on account of its very blue featureless continuum. It was previously classified as an O-type subdwarf by \citet{wegner1988}, and does differ from the SDSS quasars in that it has a bluer $(g - z)$ color. Similarly, we find that the predicted colors for quasars with $1 \lesssim z \lesssim 2$ based on the \citet{vandenberk2001} composite SDSS quasar spectrum are consistent with those of the other quasars lending further support to a non-QSO identity for this particular object. In Figure \ref{fig-colors} we also mark stars with SDSS spectra using large green star symbols. The increased number of such objects around the lower end of the stellar locus are F stars, whose spectra are collected by the SDSS expressly for the purpose of spectrophotometric calibration.


Three stars with SDSS spectra lie off the main stellar locus, and we now comment briefly on these sources. First, the star with very blue colors (lower left on each plot) has an SDSS spectrum of a high-temperature blackbody ($T_{eff} \gtrsim 40,000$~K) with slight Balmer absorption features, and has to be either a very hot O star or white dwarf. Its colors based on SDSS filters place it at the extreme end of the main sequence based on both comparison to known stars and to stellar models \citep{lenz1998,covey2007}, and it is in the \citet{bianchi2011} catalog of potential hot white dwarfs. In this latter catalog, its $(FUV - NUV)$ and $(NUV - r)$ colors indicate $T_{eff} \gtrsim 40,000$~K which matches the SDSS spectrum and full photometry from {\em GALEX}, XMM-OM, and the SDSS. We find at most one other viable candidate for this type of object in our full matched sample of XMM-OM detections with SDSS counterparts. It is 3$^{\prime\prime}$ from another faint blue star, and thus its {\em GALEX} photometry do not aid in its further characterization. The star lying below the stellar locus has the spectrum of an A-type star with strong Balmer absorption. Its location in the color-color diagram is expected for A stars, but also for blue horizontal branch stars and $z\gtrsim 3$ quasars \citep[e.g.,][]{newberg1999,yanny2000}. Finally, the reddest star with an SDSS spectrum has colors consistent with those of some observed M dwarfs, and with models for cool stars ($T_{eff} \lesssim 4000$~K) with low metallicity \citep{lenz1998,covey2007}. Its spectrum appears to be that of a carbon star, a designation that is also consistent with its SDSS colors \citep{downes2004}.

\section{Summary}\label{sec-conclude}

The archive of XMM observations that have targeted galaxy clusters is large, and for many of these observations the XMM-OM was also collecting data in UV filters. By taking the intersection of such cluster observations with the coverage of the SDSS and GALEX surveys and imposing a redshift cutoff of $z=0.5$, we have isolated a sample of 11 nearby Abell clusters with XMM-OM data in either of the $UVM2$ or $UVW1$ filters. We matched the astrometry of these images to that of the SDSS survey, made corrections for degradation of the XMM-OM system as a function of time, and stacked observations in fields that were observed in multiple campaigns. The accuracy of our image manipulation and photometry procedures were confirmed by their consistency with the OMCat and GALEX catalogs. In general, the $UVM2$ data are of comparable depth to the $NUV$ data of the GALEX AIS and the $UVW1$ data are comparable to the $u$ data of the SDSS.

After folding in available redshift data, we arrived at a sample of 726 galaxies for use in investigating the potential value of XMM-OM UV photometry for galaxy evolution studies. Of these, 520 belong to the 11 Abell clusters of the sample. The benefits of CMDs created using GALEX $NUV$ data are likely paralleled by XMM-OM $UVM2$ data, although our sample does not provide large numbers of such objects to test this claim. However, CMDs constructed using the $UVW1$ filter and SDSS $r$ show good promise for galaxy evolution studies. They show a strong red sequence with a wider real dispersion than CMDs created using SDSS $u$ and $r$ data, likely reflecting variations in star formation history among red sequence galaxies. They also provide greater separation between red and blue sequences than those obtained using only the SDSS data. Our sample is dominated by red sequence objects and we are unable to determine whether $UVW1$ data might prove useful in identifying green valley objects, although the aforementioned characteristics of $UVW1$-incorporating CMDs provides a basis for some optimism on this subject. The \citet{blanton2007} \kcorrect{} package may also be applied to investigate star formation histories and produce estimates of galaxy stellar masses that include the contributions of younger stellar populations. Finally, color-color diagrams using $UVW1$ data show promise for the identification of objects ranging from cool to hot stars, and extragalactic sources such as quasars.

\acknowledgments
NAM gratefully acknowledges the support for this work, which was provided by NASA ADP grant NNX09AC76G. We thank Antonio Talavera for his excellent work with XMM-OM calibration and the confirmation of an exposure time error for A2063, and we thank the anonymous referee for insightful comments which have improved the analysis and discussion. This research has made use of data obtained from the High Energy Astrophysics Science Archive Research Center (HEASARC) provided by NASA's Goddard Space Flight Center, and the NASA/IPAC Extragalactic Database (NED) which is operated by the Jet Propulsion Laboratory, California Institute of Technology, under contract with the National Aeronautics and Space Administration. Based on observations obtained with XMM-Newton, an ESA science mission with instruments and contributions directly funded by ESA Member States and NASA. Funding for the Sloan Digital Sky Survey (SDSS) has been provided by the Alfred P. Sloan Foundation, the Participating Institutions, the National Aeronautics and Space Administration, the National Science Foundation, the U.S. Department of Energy, the Japanese Monbukagakusho, and the Max Planck Society. The SDSS Web site is http://www.sdss.org/. The SDSS is managed by the Astrophysical Research Consortium (ARC) for the Participating Institutions. The Participating Institutions are The University of Chicago, Fermilab, the Institute for Advanced Study, the Japan Participation Group, The Johns Hopkins University, the Korean Scientist Group, Los Alamos National Laboratory, the Max-Planck-Institute for Astronomy (MPIA), the Max-Planck-Institute for Astrophysics (MPA), New Mexico State University, University of Pittsburgh, University of Portsmouth, Princeton University, the United States Naval Observatory, and the University of Washington.

\clearpage

\clearpage

\begin{deluxetable}{l r r r r l r}
\tablecolumns{7}
\tablecaption{Sample and XMM-OM Data\label{tbl-sample}}
\tablewidth{0pt}
\tablehead{
\colhead{Cluster} & \colhead{RA} & \colhead{Dec} & \colhead{z} & \colhead{XMM} & 
\colhead{Filter} & \colhead{Observe Date} \\
\colhead{} & \colhead{(J2000)} &  \colhead{(J2000)} & \colhead{} & \colhead{ObsID} & 
\colhead{} & \colhead{}
}
\startdata
Abell  119 & 00:56:21.4 & -01:15:47 & 0.0442 & 0402190501 & UVM2   & Jun 16, 2006 \\
           &            &           &        & 0012440101 & UVW1   & Jan 15, 2001 \\
Abell  400 & 02:57:38.6 & +06:02:00 & 0.0244 & 0300210501 & UVM2   & Jul 22, 2005 \\
           &            &           &        & 0404010101 & UVM2   & Aug  6, 2006 \\
Abell 1656 & 12:59:48.7 & +27:58:50 & 0.0231 & 0300530101 & UVM2   & Jun 19, 2005 \\
           &            &           &        & 0300530201 & UVM2   & Jun 17, 2005 \\
           &            &           &        & 0300530301 & UVM2   & Jun 11, 2005 \\
           &            &           &        & 0300530401 & UVM2   & Jun  9, 2005 \\
           &            &           &        & 0300530501 & UVM2   & Jun  9, 2005 \\
           &            &           &        & 0300530601 & UVM2   & Jun  7, 2005 \\
           &            &           &        & 0300530701 & UVM2   & Jun  7, 2005 \\
           &            &           &        & 0300530101 & UVW1   & Jun 18, 2005 \\
           &            &           &        & 0300530201 & UVW1   & Jun 17, 2005 \\
           &            &           &        & 0300530301 & UVW1   & Jun 11, 2005 \\
           &            &           &        & 0300530401 & UVW1   & Jun  9, 2005 \\
           &            &           &        & 0300530501 & UVW1   & Jun  8, 2005 \\
           &            &           &        & 0300530601 & UVW1   & Jun  7, 2005 \\
           &            &           &        & 0300530701 & UVW1   & Jun  6, 2005 \\
           &            &           &        & 0124711401 & UVW1   & May 29, 2000 \\
Abell 1983 & 14:52:44.0 & +16:44:46 & 0.0436 & 0091140201 & UVW1   & Feb 14, 2002 \\
Abell 2052 & 15:16:45.5 & +07:00:01 & 0.0355 & 0109920101 & UVW1   & Aug 21, 2000 \\
           &            &           &        & 0109920201 & UVW1   & Aug 21, 2000 \\
           &            &           &        & 0109920301 & UVW1   & Aug 21, 2000 \\
Abell 2063 & 15:23:01.8 & +08:38:22 & 0.0349 & 0200120401 & UVM2\tablenotemark{a} & Feb 17, 2005 \\
           &            &           &        & 0200120401 & UVW1\tablenotemark{a} & Feb 17, 2005 \\
Abell 2147 & 16:02:17.2 & +15:53:43 & 0.0350 & 0300350401 & UVM2\tablenotemark{a} & Feb  4, 2006 \\
           &            &           &        & 0300350301 & UVW1\tablenotemark{a} & Feb  2, 2006 \\
Abell 2151 & 16:05:15.0 & +17:44:55 & 0.0366 & 0147210201 & UVW1   & Aug  9, 2003 \\
Abell 2197 & 16:28:10.4 & +40:54:26 & 0.0308 & 0203710101 & UVM2   & Sep 23, 2004 \\
           &            &           &        & 0203710101 & UVW1   & Sep 23, 2004 \\
Abell 2199 & 16:28:38.5 & +39:33:06 & 0.0302 & 0008030601 & UVW1   & Aug 15, 2002 \\
           &            &           &        & 0008030301 & UVW1   & Jul  6, 2002 \\
Abell 2589 & 23:24:00.5 & +16:49:29 & 0.0414 & 0204180101 & UVM2   & Jun  4, 2004 \\
           &            &           &        & 0204180101 & UVW1   & Jun  4, 2004 \\
\enddata

\tablenotetext{a}{``Engineering'' mode rather than standard mosaic mode.}

\tablecomments{XMM observations are grouped by filter, with UVM2 listed first followed by UVW1.}

\end{deluxetable}

\clearpage

\begin{deluxetable}{l r r r r r r r r}
\tablecolumns{9}
\tablecaption{XMM-OM Selected Objects\label{tbl-nums}}
\tablewidth{0pt}
\tablehead{
\colhead{Cluster} & \colhead{$t_{int}^{UVM2}$} & \colhead{$t_{int}^{UVW1}$} & \colhead{$N_{UVM2}$} &
\colhead{$N_{UVW1}$} & \colhead{$N_{tot}$} & \colhead{$N_z$} & \colhead{$N_{gal}$} & \colhead{$N_{clus}$}\\
\colhead{} & \colhead{[ksec]} & \colhead{[ksec]} & \colhead{} & \colhead{} & \colhead{} &
\colhead{} & \colhead{} & \colhead{}
}
\startdata
Abell 119  &     3.6 &     0.8 & 28\tablenotemark{a} & 52\tablenotemark{a} & 64\tablenotemark{a} & 23\tablenotemark{a} & 21\tablenotemark{a} & 20 \\
Abell 400  &     8.1 & \nodata &      58 & \nodata &   58 &  12 &   6 &   6 \\
Abell 1656 &    12.0 &    14.6 &     201 &     632 &  652 & 340 & 333 & 161 \\
Abell 1983 & \nodata &     1.2 & \nodata &     220 &  220 &  44 &  44 &  38 \\
Abell 2052 & \nodata &     3.7 & \nodata &     229 &  229 &  57 &  38 &  35 \\
Abell 2063 & 4.0\tablenotemark{b} & 3.5 & 107 & 340 &  355 &  51 &  49 & 45 \\
Abell 2147 &     5.0 &     4.8 &     113 &     408 &  423 &  67 &  65 &  55 \\
Abell 2151 & \nodata &     2.2 & \nodata &     312 &  312 &  58 &  58 &  55 \\
Abell 2197 &     2.6 &     2.2 &      85 &     267 &  275 &  27 &  26 &  23 \\
Abell 2199 & \nodata &     3.7 & \nodata &     438 &  438 &  61 &  58 &  52 \\
Abell 2589 &     2.4 &     2.4 &      54 &     208 &  215 &  28 &  28 &  28 \\
\hline
Total      & \nodata & \nodata &     646 &    3106 & 3311 & 768 & 726 & 518 \\
\enddata   

\tablenotetext{a}{Only about half of the XMM-OM field is covered by the SDSS.}

\tablenotetext{b}{Corrected integration time. The image header indicated 8348 seconds, but obvious issues with photometry showed this to be in error. The proper integration time is 3976 seconds. See discussion in Section \ref{sec-calcheck}.}

\tablecomments{Columns: 1) cluster name, 2) approximate net integration time in kiloseconds for UVM2 data, 3) approximate net integration time in kiloseconds for UVW1 data, 4) number of UVM2-detected objects with SDSS counterparts, 5) number of UVW1-detected objects with SDSS counterparts, 6) total number of XMM-OM-detected objects (either UVM2 or UVW1) with SDSS counterparts, 7) number of objects from column 6 with redshift data, 8) number of objects from column 6 with redshifts satisfying $0.001 < z < 0.999$, and 9) number of objects roughly consistent with membership in that cluster.}

\end{deluxetable}

\begin{deluxetable}{l r r r r r r r r r r r r}
\tablecolumns{13}
\tablecaption{Measured Magnitudes Compared to \kcorrect Predictions\label{tbl-ecal}}
\tablewidth{0pt}
\tablehead{
\colhead{} & \multicolumn{3}{c}{All Galaxies} & \multicolumn{3}{c}{Elliptical} &
\multicolumn{3}{c}{Spiral} & \multicolumn{3}{c}{Uncertain} \\
\colhead{Filter} & \colhead{Mean} & \colhead{Dev} & \colhead{N} &
\colhead{Mean} & \colhead{Dev} & \colhead{N} &
\colhead{Mean} & \colhead{Dev} & \colhead{N}
}
\startdata
FUV  & -0.127 & 0.406 & 247 & -0.461 & 0.511 & 34 &  0.022 & 0.186 & 15 & -0.044 & 0.408 & 51 \\
NUV  & -0.132 & 0.298 & 242 & -0.250 & 0.306 & 33 & -0.127 & 0.126 & 15 & -0.105 & 0.319 & 50 \\
UVM2 & -0.186 & 0.365 & 122 & -0.248 & 0.350 & 19 & -0.112 & 0.284 &  8 &  0.008 & 0.348 & 21 \\
UVW1 & -0.076 & 0.278 & 232 &  0.049 & 0.159 & 33 & -0.312 & 0.173 & 14 & -0.034 & 0.266 & 48 \\
u    &  0.085 & 0.090 & 245 &  0.119 & 0.065 & 34 &  0.048 & 0.053 & 15 &  0.077 & 0.062 & 50 \\
g    & -0.016 & 0.064 & 248 & -0.005 & 0.015 & 34 & -0.003 & 0.021 & 15 & -0.007 & 0.018 & 49 \\
r    & -0.007 & 0.056 & 248 &  0.005 & 0.010 & 34 & -0.002 & 0.004 & 14 &  0.004 & 0.013 & 51 \\
i    & -0.009 & 0.046 & 246 & -0.009 & 0.008 & 34 & -0.001 & 0.009 & 15 & -0.009 & 0.012 & 48 \\
z    &  0.004 & 0.010 & 245 &  0.002 & 0.002 & 34 &  0.001 & 0.002 & 15 &  0.002 & 0.004 & 50 \\
\enddata

\tablecomments{Values are for the difference between the measured magnitudes and those predicted based on the GALEX and SDSS photometry input to \kcorrect . A round of 2$\sigma$ clipping was used to remove strong outliers before computing the mean and deviation. Morphologies are taken from Table 2 of \citet{lintott2011}, and represent ``clean'' assignments after accounting for known biases -- hence the relatively large fraction of ``Uncertain'' morphologies.}

\end{deluxetable}

\begin{deluxetable}{r r r r r r r r r r r r r}
\tablecolumns{13}
\tablecaption{Accuracy of Photometric Redshifts\label{tbl-photoz}}
\tabletypesize{\scriptsize}
\tablewidth{0pt}
\tablehead{
\colhead{} & \multicolumn{4}{c}{All Galaxies} & \multicolumn{4}{c}{Cluster Galaxies} &
\multicolumn{4}{c}{Background Galaxies} \\
\colhead{Filters Used} & \colhead{$N_{full}$} & \colhead{$\sigma_{z,full}$} &
\colhead{$N_{clip}$} & \colhead{$\sigma_{z,clip}$} &
\colhead{$N_{full}$} & \colhead{$\sigma_{z,full}$} &
\colhead{$N_{clip}$} & \colhead{$\sigma_{z,clip}$} &
\colhead{$N_{full}$} & \colhead{$\sigma_{z,full}$} &
\colhead{$N_{clip}$} & \colhead{$\sigma_{z,clip}$} 
}
\startdata
$u~g~r~i~z$           & 726 & 0.131 & 711 & 0.078 & 518 & 0.128 & 506 & 0.055 & 205 & 0.121 & 202 & 0.103 \\
$F~N~u~g~r~i~z$       & 239 & 0.147 & 233 & 0.072 & 161 & 0.163 & 155 & 0.053 &  77 & 0.099 &  74 & 0.076 \\
$F~u~g~r~i~z$         & 250 & 0.149 & 243 & 0.067 & 167 & 0.159 & 161 & 0.048 &  82 & 0.124 &  81 & 0.091 \\
$N~u~g~r~i~z$         & 497 & 0.126 & 486 & 0.065 & 360 & 0.124 & 353 & 0.043 & 135 & 0.120 & 131 & 0.097 \\
$UVM2~UVW1~u~g~r~i~z$ & 174 & 0.136 & 170 & 0.086 &  97 & 0.133 &  95 & 0.048 &  75 & 0.134 &  73 & 0.110 \\
$UVM2~u~g~r~i~z$      & 197 & 0.138 & 192 & 0.087 & 114 & 0.132 & 112 & 0.054 &  81 & 0.143 &  80 & 0.133 \\
$UVW1~u~g~r~i~z$      & 698 & 0.136 & 679 & 0.079 & 499 & 0.130 & 485 & 0.053 & 196 & 0.139 & 190 & 0.110 \\
$UVW1~g~r~i~z$        & 698 & 0.137 & 681 & 0.082 & 499 & 0.130 & 485 & 0.050 & 196 & 0.146 & 192 & 0.126 \\
\enddata

\end{deluxetable}

\end{document}